# Exciton-dominated Dielectric Function of Atomically Thin MoS$_2$ Films

Yiling Yu[2], Yifei Yu[1], Yongqing Cai[3], Wei Li[4], Alper Gurarslan[1,5], Hartwin Peelaers[6]
David E. Aspnes[2], Chris G. Van de Walle[6], Nhan V. Nguyen[4], Yong-Wei Zhang[3], Linyou Cao[1,2]*

[1]Department of Materials Science and Engineering, North Carolina State University, Raleigh, NC 27695, USA; [2]Department of Physics, North Carolina State University, Raleigh NC 27695, USA; [3] Institute of High Performance Computing, A*STAR, Singapore 138632; [4]Semiconductor and Dimensional Metrology Division, National Institute of Standards and Technology, Gaithersburg, Maryland 20899, USA; [5]Department of Fiber and Polymer Science, North Carolina State University, Raleigh, NC 27695, USA; [6]Materials Department, University of California, Santa Barbara, CA 93106, USA

**Abstract**

We systematically measure the dielectric function of atomically thin MoS$_2$ films with different layer numbers and demonstrate that excitonic effects play a dominant role in the dielectric function when the films are less than 5-7 layers thick. The dielectric function shows an anomalous dependence on the layer number. It decreases with the layer number increasing when the films are less than 5-7 layers thick but turns to increase with the layer number for thicker films. We show that this is because the excitonic effect is very strong in the thin MoS$_2$ films and its contribution to the dielectric function may dominate over the contribution of the band structure. We also extract the value of layer-dependent exciton binding energy and Bohr radius in the films by fitting the experimental results with an intuitive model. The dominance of excitonic effects is in stark contrast with what reported at conventional materials whose dielectric functions are usually dictated by band structures. The knowledge of the dielectric function may enable capabilities to engineer the light-matter interactions of atomically thin MoS$_2$ films for the development of novel photonic devices, such as metamaterials, waveguides, light absorbers, and light emitters.

* To whom correspondence should be addressed.
Email: lcao2@ncsu.edu

Two-dimensional (2D) transition metal dichalcogenide (TMDC) materials have been known exhibiting strong exciton binding energy that may be one order of magnitude larger than conventional semiconductor materials[1-6]. However, how the extraordinarily strong exciton binding energy could affect the light-matter interactions such as dielectric functions of the materials has remained unexplored. The lack of knowledge about the dielectric function has significantly limited the application of 2D TMDC materials in many exciting photonic fields such as metamaterials[7], which relies on the sophisticated manipulation of effective dielectric functions to enable novel optical functionalities. In this work we have measured the dielectric function of atomically thin $MoS_2$ films and discovered that it is dominated by the effect of the tightly bound excitons, as evidenced by an anomalous dependence of the dielectric function on the layer number. The dielectric function decreases with the layer number increasing when the $MoS_2$ films are less than 5 layers thick, but turn to increase with the layer number for thicker. We also quantitatively evaluate the exciton binding energy and Bohr radius of the thin films by fitting the experimental results with an intuitive model. The observed dominance of excitonic effects in the dielectric function is in stark contrast with what expected at conventional materials, whose dielectric functions are usually dictated by band structures[8, 9]. Our success in this discovery is built upon a unique self-limiting chemical vapor deposition (CVD) process that we have recently developed[10]. The self-limiting CVD process can be used to grow centimeter-scale, uniform, and high quality atomically thin $MoS_2$ films with controlled layer numbers and remarkable uniformity (Fig. S1-S5). This allows us to examine the dielectric function of $MoS_2$ films as a function of well-defined layer numbers. Our work is different from earlier research for the dielectric function of $MoS_2$ films[11, 12], whose results are likely inaccurate due to the lack of satisfactory uniformity or precise control of the layer number.

We measured the dielectric function ($\varepsilon_1 + i\varepsilon_2$) of as-grown $MoS_2$ films on sapphire substrates using spectroscopic ellipsometry[13]. Fig. 1a-b shows the real $\varepsilon_1$ and imaginary $\varepsilon_2$ parts of the dielectric function in the visible range that are derived from experimental measurements (see Fig. S7-S8 for the fit between experimental and simulated results). Owing to the extreme geometrical anisotropy of the film, what we obtained is actually the in-plane component of the dielectric tensor because the out-of-plane dielectric function may only contribute trivially to the optical response due to the difficulty in exciting the vertical dipole of the atomically thin film[13]. As further evidence for the measured in-plane dielectric function, we performed the spectroscopic ellipsometry at different incident angles (40°-75°), and all of them ended up with giving very similar dielectric functions. The dielectric function of bulk $MoS_2$ is also measured and plotted in Fig. 1a-b as a reference, the result of which is consistent with what reported previously[14]. The three peaks in the spectral dielectric function can be assigned to *A*, *B*, and *C* from low to high energies, respectively[15-18]. The *A* and *B* peaks are related with the transition from the spin-orbit split valence bands to the lowest conduction band at the *K* and *K'* points, while the *C* peak is associated with the transition from the valence band to the conduction band at the part of the Brillouin zone between the $\Lambda$ and $\Gamma$ point[16, 17].

The measured dielectric function is not sensitive to the synthetic process or the substrate. The dielectric functions measured from the $MoS_2$ grown by using $MoCl_5$ and S as the precursors[10] and by using $MoO_3$ and S as the precursors[19] are essentially identical (Fig. S9a). We also find that the dielectric functions of the as-grown $MoS_2$ films on sapphire substrates and those transferred onto $SiO_2$/Si substrates are identical (Fig. S9b). Additionally, the dielectric function of the film is stable under ambient environment. We monitored the dielectric function of the as-grown $MoS_2$ films on sapphire substrates as a function of the time for the films to be exposed to ambient environment. We monitored the dielectric function of the films exposed to ambient environment for more than one week and found no change in the measured result (Fig. S10). The result we measured for the monolayer $MoS_2$ film is consistent with what previously measured using spectroscopic ellipsometry[20] but is 10-15% less than the results derived from absorption spectra[21]. We do like to point out some difference in the spectroscopic ellipsometry used by us as well as Ref. 20 and the spectroscopic absorption technique used in Ref. 21.

Spectroscopic ellipsometry is the most established technique for the measurement of dielectric functions, in which two parameters are measured at each wavelength and the dielectric function can be uniquely determined in any spectrum range with the thickness information of the film indepedently determined by AFM. The dielectric function may be derived from spectroscopic reflection using the Kramer-Kronig relationship as well. But to precisely find out the dielectric function using the Kramer-Kronig relationship requires information of the absorption in the entire spectral range. In Ref. 21 the absorption of the monolayer in the range higher than 3 eV, whose value is not experimentally available, is assumed to be equal to that of the bulk counterpart. This assumption might overestimate the dielectric function to some degree. We believe this is likely the reason why our result is around 10-15% less than the result reported in Ref.21.

Significantly, the measured dielectric function shows an anomalous dependence on the layer number (Fig. 1a-b). It decreases with the layer number increasing when the film is less than 5-7 layers thick and then turns to increase with the layer for thicker films. We were very careful to ensure no artifact introduced in the measurement. More specifically, we performed extensive AFM for each of the films studied prior to the ellipsometry measurement and confirmed the atomic-scale smoothness (roughness usually < 0.5 nm except the 8L, 9L and 10L films, whose roughness is a little bit larger in the range of 0.7-0.9 nm, see Fig. S1-S5) and excellent uniformity of the film. Additionally, for the result of each layer number, we measured at least three different sets of samples and observed only minor variation (5%) in the resulting dielectric function. To further illustrate this anomalous layer-dependence, we extract the imaginary part of the dielectric functions at the *A*, *B*, and *C* peaks from Fig. 1b and plot it as a function of the layer number (Fig.1c). The result clearly shows a decrease and then an increase in the dielectric function with the layer number continuously increasing from one. The layer dependence is similar for all the *A*, *B*, and *C* peaks (Fig. 1d). Given the similarity in the layer dependence, we only focus on the *C* peak in the following discussion.

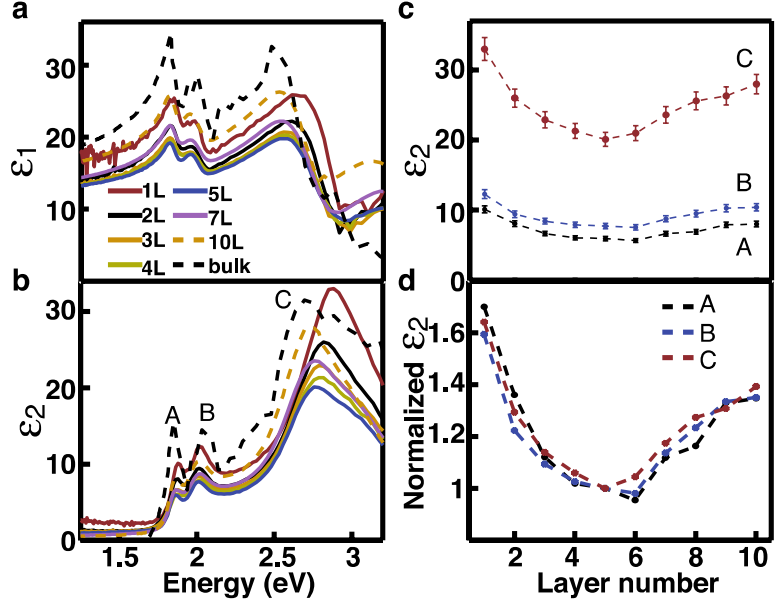

**Figure 1. Anomalous layer-dependence of the dielectric function of 2D MoS$_2$.** (a-b) Real and imaginary parts of the dielectric function of 2D MoS$_2$ vs. layer number. Also given is the dielectric function of bulk MoS$_2$. The three peaks can be assigned to *A*, *B*, and *C* excitons as labeled. Corresponding refractive indexes of the films are given in Figure S6 and Table S1 of the Supplementary Information. (c) The dependence of the imaginary part $\varepsilon_2$ of the dielectric function at the *A*, *B*, and *C* peaks on layer number. The error bar is 5% and estimated from the measurement results of multiple samples. (d) Normalized $\varepsilon_2$ at the *A*, *B*, and *C* peaks vs. layer number. The normalization is performed with respect to the corresponding value of each peak in the 5-layer MoS$_2$. Error bar is ignored for visual convenience.

To obtain physical insights into the observed layer dependence, we examine the dielectric function from the perspective of quantum mechanics. We only focus on the imaginary part $\varepsilon_2$ because the real part $\varepsilon_1$ can be deterministically correlated to $\varepsilon_2$ by the well-established Kramer-Kronig equation[22]. Fundamentally, $\varepsilon_2$ is related with interband transitions as[8, 22, 23]

$$\varepsilon_2(\omega) = \frac{4\pi^2 e^2}{m_0^2 \omega^2} J_{cv} |p_{cv}|^2 |U(0)|^2 \frac{\Gamma/2}{(E_{cv} - \hbar\omega)^2 + (\Gamma/2)^2} \quad (1)$$

where $\omega$ is the frequency, $\hbar$ is the Planck's constant, $e$ and $m_0$ are the charge and mass of free electrons, $J_{cv}$ is the joint density of the initial (valence band) and final (conduction band) states involved in the transition. $p_{cv}$ is an optical matrix element indicating the probability of the transition from the initial to final states. It consists of an integral over a unit cell that involves the momentum operator as well as the unit cell wavefunctions in the conduction and valence bands. $|U(0)|^2$ represents the effect of excitons on the oscillator strength of the interband transition, where $U$ is the relative motion wavefunction of the eletrons and holes bound by Coulomb interactions and $0$ indicates the physical overlap of the electron and hole wavefunctions. $E_{cv}$ is the optical energy gap between the conduction and valence bands involved and $\Gamma$ is a damping constant determining the bandwidth of the interband transition.

For simplicity, we only focus on the dielectric function at the peak position (on-resonance) as shown in Fig. 1c-d, where $E_{cv} - \hbar\omega = 0$ and eq.(1) can be simplified as

$$\varepsilon_2(\omega) = \frac{4\pi^2 e^2}{m_0^2 \omega^2} J_{cv} |p_{cv}|^2 |U(0)|^2 \frac{2}{\Gamma} \qquad (2)$$

Physically, $\Gamma$ represents the width of the peak. From Fig. 1b the peak can be found phenomenally remaining to be similar in the films with different layer numbers. Therefore, it is reasonable to consider that $\Gamma$ is independent of the layer number. The optical matrix element $p_{cv}$ is also independent of the layer number because it is only related with unit cells and unit cell wavefunctions, both of which are not dependent on the layer number. The independence of $p_{cv}$ on geometrical features has previously been demonstrated at quantum wells[24]. Therefore, eq. (2) can be further simplified as

$$\varepsilon_2(\omega) = A_0 J_{cv} |U(0)|^2 \qquad (3)$$

where $A_0$ includes all the terms independent of the layer number. Eq. (3) indicates that the layer dependence of the dielectric function may result from only two parameters: the joint density of states $J_{cv}$ and the excitonic effect $|U(0)|^2$.

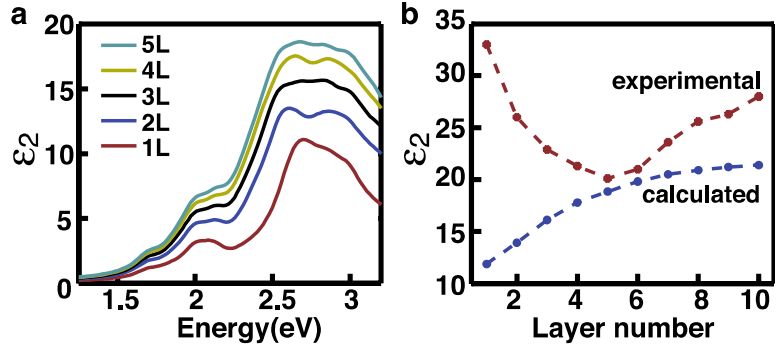

**Figure 2. Comparison of the measured and calculated dielectric function of 2D MoS$_2$.** (a) Calculated imaginary part $\varepsilon_2$ of the dielectric function of MoS$_2$ with different thickness. (b) Comparison of experimental and calculated results for $\varepsilon_2$ at the *C* peak as a function of the layer number. The error bar in the experimental result is ignored for visual convenience.

We can get more insight by further examining the specific layer dependence of the two parameters $J_{cv}$ and $|U(0)|^2$. The joint density of states $J_{cv}$ is determined by the band structure and expected to monotonically increase with the layer number. This is because the density of states in 2D materials is well known to be smaller than that in 3D materials[8, 22] and the increase of the layer number in effect enables a continuous evolution from two dimensions (monolayers) to three dimensions (bulk). To quantitatively elucidate the layer dependence of $J_{cv}$, we calculate the $\varepsilon_2$ of MoS$_2$ films using density functional theory (DFT) techniques without considering excitonic effects[18]. The calculation result is given in Fig. 2a and essentially represents the density of states $J_{cv}$. It reproduces the major spectral features of the experimental results and indeed shows a monotonic increase with the layer number. For the convenience of comparison, we plot the calculated and measured dielectric functions at the *C* peak as a function of the layer number in Fig. 2b. The calculated results are understandably smaller than the experimental results due to the exclusion of excitonic effects. Of our interest is to compare the layer dependence in the calculated and experimental results. The similarity of the two results for the films > 5L suggests that the observed layer-dependent increase in the dielectric function in the relatively thicker films may be correlated to the effect of the density of states $J_{cv}$. However, for the films less than 5 layers thick the calculated layer-dependence is opposite to the experimental observation (Fig.2b). This can be correlated to the other parameter not considered in the calculation, the excitonic effect $|U(0)|^2$. The excitonic effect is expected to be strong and to quickly decrease with the layer number due to the well known layer-dependent exponential decrease of exciton binding energy[15].

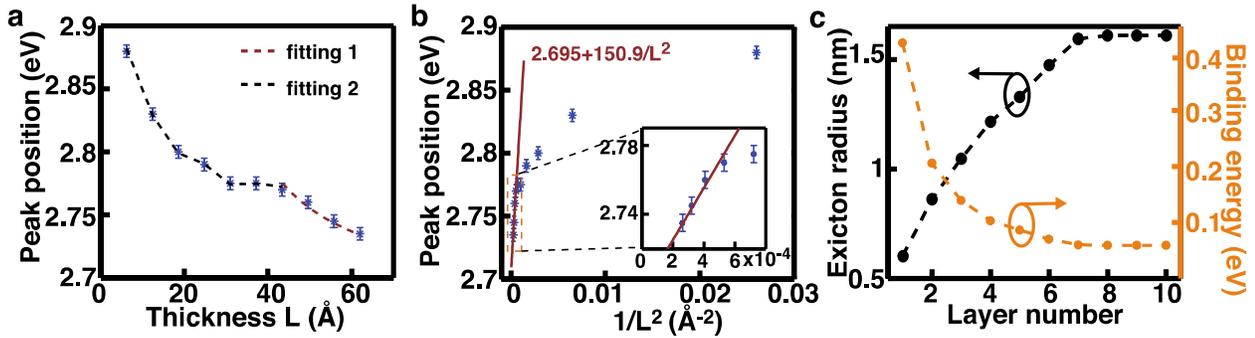

**Figure 3. Strong, layer-dependent excitonic effects in atomically thin MoS$_2$ films.** (a) The position of the *C* peak in MoS$_2$ films as a function of the thickness of the film *L*. The error bar ±0.005 eV results from the possible errors in determining the peak position. The dashed lines are the fitting results using the model of infinite quantum wells (fitting 1) and the quantum well in fraction space (fitting 2) (b) The position of the *C* peak in MoS$_2$ films as a function of $1/L^2$, where *L* is the thickness of the film. The red line is the fitting results using the model of infinite quantum well with the fitting equation given as shown. The inset is a magnified version of the area indicated by the dashed yellow rectangle. (c) The dependence of the binding energy and exciton radius in MoS$_2$ films on the layer number.

The layer-dependent excitonic effect can be understood more quantitatively by examining the excitonic peak position with an intuitive model that involves quantum confinement and exciton binding energy. Of our interest is to quantitatively evaluate the layer dependence of the exciton binding energy from the observed evolution of the *C* exciton peak position with the layer number (Fig. 3a). The excitonic peak position (i.e. optical bandgap) is equal to the electronic bandgap minus the exciton binding energy, and its layer dependence originates from the layer dependence of both components. Should the layer dependence of the electronic bandgap be found out, we would be able to figure out the layer dependence of the exciton binding energy from the measured excitonic peak position. It is actually very difficult to experimentally or theoretically evaluate the electronic bandgap of MoS$_2$ with different layer numbers. However, we find out that the layer dependence of the electronic bandgap may be reasonably correlated to quantum confinement effects. The peak position of the $C$ exciton $E_C$ in the films thicker than 7 layers shows a linear dependence on $1/L^2$ (Fig. 3b), and can be fitted by the conventional model of infinite quantum wells as $E_C = E_g + R_y + \pi^2\hbar^2/2m_{eff}L^2 - R_y$. where $E_g$ is the position (optical bandgap) of the $C$ peak in bulk MoS$_2$ materials ($E_g$ = 2.695 eV as measured in Fig.1b), $R_y$ is the exciton binding energy and assumed not changing with the thickness in the model of infinite quantum wells, and $m_{eff}$ is the reduced electron-hole effective mass of the film. The introduction of $R_y$ in the equation is to illustrate that the optical bandgap is equal to the electronic bandgap (the first three terms) minus the exciton binding energy (the last term). The fitting to the experimental results (the red lines in Fig. 3a-b) gives $E_C$ = 2.695 eV + 150.9/$L^2$, from which we can derive the reduced effective mass $m_{eff}$ = 0.250 $m_0$ for the $C$ extions in bulk MoS$_2$ and the films thicker than 7 layers. We can also derive the exciton binding energy $R_y$ = 58.9 meV and Bohr radius $a_b$ = 1.61 nm in bulk MoS$_2$ and thick films, in which the static dielectric constant is set to be 7.6 as measured for bulk MoS$_2$ previously[25].

The peak position $E_C$ in the films thinner than 5-7 layers shows apparent deviation from the model of infinite quantum wells (Fig. 3b). Instead, we can fit the experimental results with a model of quantum wells in fractional dimensional space as[26, 27]

$$E_C = E_g + R_y + \frac{\pi^2\hbar^2}{2m_{eff}L^2}\left[(D-1)/2\right]^2 - \frac{R_y}{\left[(D-1)/2\right]^2} \quad (4)$$

where $D$ is the effective dimensionality that is defined by the ratio of the exciton binding energy $R_y^*$ in the films and that of bulk MoS$_2$ $R_y$ as $[(D-1)/2]^2 = R_y/R_y^*$. Again, the first three terms of eq. (4)

represent the electronic bandgap and the last term indicates the exciton binding energy. The factor of $[(D-1)/2]^2$ in the third term originates from the change in the effective mass associated with the effective dimensionality. For the films thicker than 7 layers, $D$ is 3 and eq. (4) is then reduced to the equation for infinite quantum wells. By fitting the experimental results with eq. (4), we can have the effective dimensionality $D$ as 1.75, 2.07, 2.30, 2.51, 2.65, and 2.83 for the films in layer number of 1, 2, 3, 4, 5, and 6, respectively. We can then derive the corresponding exciton binding energies using $R_y^* = R_y/[(D-1)/2]^2$ as 0.421eV, 0.206 eV, 0.139 eV, 0.103 eV, 0.0865 eV, and 0.0704 eV; we can also derive the corresponding Bohr radius of excitons from $a_b^* = a_b(D-1)/2$ [26, 27] as 0.602 nm, 0.861 nm, 1.04 nm, 1.22 nm, 1.33 nm, and 1.47 nm, respectively. These results are plotted in Fig. 3c.

The model we used to fit the experimental result is based on an assumption that the layer dependence of the electronic bandgap can be ascribed to the effect of quantum confinement. This is supported by our experimental results, in particular, the consistence between the observed peak position of the films thicker than 7 layers and what predicted from the model of infinite quantum wells. However, more theoretical and experimental studies would be necessary to provide more rigorous support, which is to our best knowledge expected to very difficult. Nevertheless, the result we obtained by fitting the experimental results using this model seems to be reasonable when compared to the limited number of studies on the exciton binding eerngy and Bohr radisu available in the literature. There is not study that would allow us to systematically crosscheck all of our results. For instance, the binding energy $R_y$ = 58.9 meV and Bohr radius $a_b$ = 1.61 we derived for the $C$ exciton in bulk MoS$_2$ and thick films is reasonably consistent with the binding energy and Bohr radius reported for the $A$ exciton in bulk MoS$_2$, which are 87.2 meV and 1.11 nm, respectively[28]. Additionally, the Bohr radius (0.602 nm) we derived for the $C$ exciton in monolayer MoS$_2$ nicely matches the theoretical prediction, ~ 0.5 nm[17]. The derived binding energy (0.421 eV) is reasonable compared with what reported for the $A$ exciton, which is believed to be 0.4-0.6 eV in monolayer $MoS_2$[18, 29].

With the information of the exciton binding energy and radius, the observed layer dependence of the dielectric function can be intuitively understood from a perspective of geometric confinement. Fig. 4 shows the comparison between the size of excitons and the thickness of the film. While the film is highly anisotropic, the size of the exciton is schematically illustrated by the diameter of a sphere anyway. This is because eq. (4), which we used to derive the exciton radius, treats the excitons as spheres in an isotropic space by converting the geometrical anisotropy into fractional dimensionality[26, 27]. The size of the exciton in bulk $MoS_2$, 3.22 nm, is close to the thickness of the 5L film, 3.10 nm (Fig. 4a). Therefore, the exciton in $MoS_2$ films is expected to start experiencing substantial geometrical confinement when the layer number of the film is decreased to 5, which may lead to decrease in the exciton size. Intuitively, a smaller exciton radius can better facilitate the spatial overlap of the electron and hole wavefunctions and subsequently cause larger amplitude in $|U(0)|^2$. The layer-dependent decrease of the dielectric function is expected when the layer-dependent decrease of the excitonic effect offsets or even exceeds the layer-dependent increase of the join density of states.

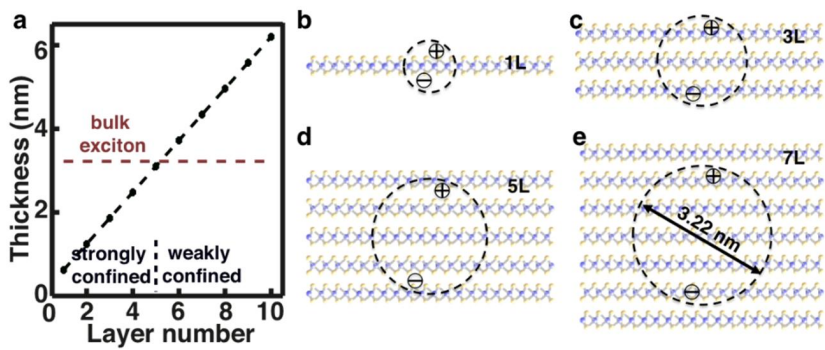

**Figure 4. Geometric confinement of excitons in $MoS_2$ films.** (a) Comparison of the size of the exciton in bulk $MoS_2$ with the thickness of $MoS_2$ films. The regime where the film is thinner than the exciton size is categorized as strong confinement. (b-e) Schematic illustration of the size of excitons in the films with different layer numbers.

The main conclusion we draw from the analysis of the *C* exciton, i.e. the dominance of excitonic effects in the dielectric function, can be applied to the *A* and *B* excitons as well due to the similar layer dependence in corresponding dielectric functions (Fig.1c-d). However, it is difficult to quantitatively extract the binding energy and Bohr radius for the *A* and *B* excitons as what we did for the *C* exciton. This is because that the positions of the *A* and *B* excitons do not show substantial layer dependence as what observed as the *C* exciton, which makes the fitting by the intuitive model difficult. One reason for the less layer dependence observed at the *A* and *B* excitons could be related with the random variation of the position of the *A* and *B* excitons due to local doping effect from the substrate, which may be as large as ~ 10 meV [30]. Another reason could be related with the better localization of the A and B excitons in the plane of the film, which may lead to a less dependence on interlayer interactions[31].

While our work mainly focuses on MoS$_2$, we believe that similar dominance of excitonic effects in the dielectric function could generally exist in all the atomically thin semiconducting TMDC materials. Our result bears significant implications for the development of photonics devices with 2D TMDC materials. The obtained dielectric function and refractive index (Table S1) for the MoS$_2$ with different layer numbers can be immediately useful for the rational design of photonic devices. In particular, as excitons are subject to influence of electric or magnetic fields, the dominance of excitonic effects in the dielectric function makes atomically MoS$_2$ films an unprecedented platform that may enable the development of field-effect photonics, whose optical functionalities would be tuned by external electric or magnetic fields.

**Acknowledgements**

This work was supported by a Young Investigator Award from the Army Research Office (W911NF-13-1-0201). Y.C. and Y.Z. gratefully acknowledge the use of computing resources at the A*STAR

Computational Resource Centre, Singapore. H.P. and C.G.V.d.W were supported by the U.S. Department of Energy, Office of Science, Basic Energy Sciences, under Award DE-FG02-07ER46434.

**Author contributions**

L.Y. conceived the idea. Yiling Yu performed most of the elliposometry measurements and other characterizations of the films, Yifei Yu synthesized the films. A. G. performed the transfer of the film and some of the AFM measurements. W. L. and N. V. N. helped with part of the ellipsometry measurements. Y. C. and Y. Z. conducted the theoretical calculations. H. P. and C. G. V. d. W also contributed to the theoretical calculations. L. Y., Yiling Yu, and D. E. A. analyzed the results. L. Y. and Yiling Yu wrote the paper. All the people were involved in reviewing the paper.

**Competing financial interests**

The authors declare no competing financial interests.

**References**

1. Cao, L.Y. Two-dimensional transition-metal dichalcogenide materials: Toward an age of atomic-scale photonics. *Mrs Bull* **40**, 592-599 (2015).
2. Mak, K.F. et al. Tightly bound trions in monolayer MoS2. *Nat Mater* **12**, 207-211 (2013).
3. Mak, K.F., Lee, C., Hone, J., Shan, J. & Heinz, T.F. Atomically Thin MoS2: A New Direct-Gap Semiconductor. *Phys Rev Lett* **105** (2010).
4. Splendiani, A. et al. Emerging Photoluminescence in Monolayer MoS2. *Nano Lett* **10**, 1271-1275 (2010).
5. Wang, Q.H., Kalantar-Zadeh, K., Kis, A., Coleman, J.N. & Strano, M.S. Electronics and optoelectronics of two-dimensional transition metal dichalcogenides. *Nat. Nanotech.* **7**, 699-712 (2012).
6. Xu, X., Yao, W., Xiao, D. & Heinz, T.F. Spin and Pseudospins in Layered Transition Metal Dichalcogenides. *Nature Physics* **10**, 343–350 (2014).
7. Smith, D.R., Pendry, J.B. & Wiltshire, M.C.K. Metamaterials and negative refractive index. *Science* **305**, 788-792 (2004).
8. Miller, D.A.B. Quantum Mechanics for Scientists and Engineer. (Cambridge University Press, New York, NY; 2008).
9. Palik, E.D. Handbook of Optical Constants of Solids. (Acadmic Press, London; 1985).
10. Yu, Y.F. et al. Controlled Scalable Synthesis of Uniform, High-Quality Monolayer and Few-layer MoS2 Films. *Sci Rep-Uk* **3** (2013).

11. Yim, C. et al. Investigation of the optical properties of MoS2 thin films using spectroscopic ellipsometry. *Appl. Phys. Lett.* **104**, 103114 (2014).
12. Liu, H.-L. et al. Optical properties of monolayer transition metal dichalcogenides probed by spectroscopic ellipsometry. *Appl. Phys. Lett.* **105**, 201905 (2014).
13. Tompkins, H.G. & Irene, E.A. Handbook of Ellipsometry. (William Andrew, Norwich, NY; 2005).
14. Beal, A.R. & Hughes, H.P. Kramers-Kronig Analysis of the Reflectivity Spectra of 2h-Mos2, 2h-Mose2 and 2h-Mote2. *J Phys C Solid State* **12**, 881-890 (1979).
15. Komsa, H.-P. & Krasheninnikov, A.V. Effects of confinement and environment on the electronic structure and exciton binding energy of MoS2 from first principles. *Phys. Rev. B* **86**, 241201 (2012).
16. Kozawa, D. et al. Photocarrier relaxation pathway in two-dimensional semiconducting transition metal dichalcogenides. *Sci. Rep.* **5**, 4543 (2014).
17. Qiu, D.Y., Jornada, F.H.d. & Louie, S.G. Optical Spectrum of MoS2: Many-Body Effects and Diversity of Exciton States. *Phys. Rev. Lett.* **111**, 216805 (2013).
18. Shi, H.L., Pan, H., Zhang, Y.W. & Yakobson, B.I. Quasiparticle band structures and optical properties of strained monolayer MoS2 and WS2. *Physical Review B* **87** (2013).
19. Lee, Y.H. et al. Synthesis of Large-Area MoS2 Atomic Layers with Chemical Vapor Deposition. *Adv Mater* **24**, 2320-2325 (2012).
20. Li, W. et al. Broad Band Optical Properties of Large Area Monolayer CVD Molybdenum Disulfide. *Phy. Rev. B* **90**, 195434 (2014).
21. Li, Y. et al. Measurement of the optical dielectric function of monolayer transition-metal dichalcogenides: MoS2, MoSe2, WS2, and WSe2. *Phys. Rev. B* **90**, 205422 (2014).
22. Chuang, S.L. Physics of Photonic Devices. (John Wiley & Sons, New York, NY; 1995).
23. Miller, D.A.B., Weiner, J.S. & Chemla, D.S. Electric-Field Dependence of Linear Optical-Properties in Quantum-Well Structures - Wave-Guide Electroabsorption and Sum-Rules. *Ieee J Quantum Elect* **22**, 1816-1830 (1986).
24. Weisbuch, C. & Vinter, B. Quantum Semiconductor Structures: Fundamentals and Applications (Academic Press, San Diego; 1991).
25. Evans, B.L. & Young, P.A. Optical absorption and dispersion in molybdenum disulphide. *Proc. R. Soc. Lond. A* **284**, 402-422 (1965).
26. He, X.F. Fractional dimensionality and fractional derivative spectra of interband optical transitions. *Phys. Rev. B* **42**, 11751-11756 (1990).
27. He, X.F. Excitons in anisotropic solids: The model of fractional-dimensional space. *Phys. Rev. B* **43**, 2063-2069 (1991).
28. Fortin, E. & Raga, F. Excitons in molybdenum disulphifle. *Phys. Rev. B* **11**, 905-912 (1965).
29. Klots, A.R. et al. Probing excitonic states in suspended two-dimensional semiconductors by photocurrent spectroscopy. *Sci. Rep.* **4**, 6608 (2014).
30. Buscema, M., Steele, G.A., Zant, H.S.J.v.d. & Castellanos-Gomez, A. The effect of the substrate on the Raman and photoluminescence emission of single layer MoS2. *Nano Res.*, DOI: 10.1007/s12274-12014-10424-12270 (2014).
31. Gurarslan, A. et al. Surface Energy-Assisted Transfer of Centimeter-Scale Monolayer and Fewlayer MoS2 Films onto Arbitrary Substrates. *ACS Nano* **8**, DOI: 10.1021/nn5057673 (2014).

*Supplementary Information*

**Exciton-dominated Dielectric Function of Atomically Thin MoS$_2$ Films**

Yiling Yu[2], Yifei Yu[1], Yongqing Cai[3], Wei Li[4], Alper Gurarslan[1,5], Hartwin Peelaers[6] David E. Aspnes[2], Chris G. Van de Walle[6], Nhan V. Nguyen[4], Yong-Wei Zhang[3], LinyouCao[1,2]*

[1]Department of Materials Science and Engineering, North Carolina State University, Raleigh, NC 27695, USA; [2]Department of Physics, North Carolina State University, Raleigh NC 27695, USA; [3]Institute of High Performance Computing, A*STAR, Singapore 138632; [4]Semiconductor and Dimensional Metrology Division, National Institute of Standards and Technology, Gaithersburg, Maryland 20899, USA; [5]Department of Fiber and Polymer Science, North Carolina State University, Raleigh, NC 27695, USA; [6]Materials Department, University of California, Santa Barbara, CA 93106, USA

* To whom correspondence should be addressed.
Email: lcao2@ncsu.edu

This PDF file includes
Methods
Fig. S1-S12
Table. S1 Tabulated refractive index of MoS$_2$ films References
S1-S4

**Methods**

*Synthesis of centimeter-scale MoS₂ films.* The centimeter monolayer and fewlayer $MoS_2$ films were grown using a chemical vapor deposition (CVD) process that we have previously developed[1]. In a typical growth, 4-20 mg molybdenum chloride ($MoCl_5$) powder (99.99%, Sigma-Aldrich) and 1g sulfur powder (Sigma-Aldrich) was placed at the upstream of a tube furnace. Receiving substrates (typically sapphire) were placed in the downstream of the tube. Other typical experimental conditions for the growth include a temperature of 850 °C, a flow rate of 50 sccm, and a pressure around 2 Torr. The layer number was controlled by controlling the amount of precursors ($MoCl_5$) used in the synthesis. We also grow centimeter-scale monolayer $MoS_2$ films using a similar chemical vapor deposition process but with $MoO_3$ and sulfur being used as precursors

The composition and structure of the synthesized $MoS_2$ films were extensively characterized by a variety of tools previously. In this work we mainly focused on characterizing the thickness and surface morphology of the synthesized films with optical microscopes and atomic force microscope (AFM, Veeco Dimension-3000).

*Ellipsometry measurements of MoS₂ films.* The ellipsometry measurement was performed with Woollam VASE (Variable Angle Spectroscopic Ellipsometer, J.A. Woollam Co.) with a Xenon light source in range of 200-1100nm. Typical incident angle was set at $65^0$, but we confirmed that the incident angle does not affect the resulting dielectric constants. The ellipsometry measurement essentially monitors changes in the polarization state of incident light and the light reflected from the films. It yields two spectral parameters ($\psi$ and $\Delta$) related with the amplitude (tan $\psi$) and phase $\Delta$ of a reflectance ratio $\rho$, which indicates the ratio of the reflection coefficients for *p*-polarized (parallel to the plane of incidence) and *s*-polarized (perpendicular to the plane of incidence) light, $\rho = r_p/r_s = (\tan\psi)e^{i\Delta}$. To retrieve the optical constant from the measured results, we perform regression fitting using the Fresnel's equations of a simple two-layer model that consists of a $MoS_2$ film on top of a semi-infinite substrate. Precise information of the thickness of the film is required for the fitting, which we have obtained by AFM measurements in experiments. We have also confirmed that the surfaces of all the films studied are atomically smooth (roughness < 1 nm) by the AFM measurements (Fig. S1-S3). The typical fitting for experimental results is given in Fig. S4.

*First-principles calculations.* Our DFT calculations are carried out using the plane wave code Vienna ab initio simulation package (VASP)[2] with the generalized gradient approximation (GGA). Spin-orbital coupling calculations using the projector augmented wave method with the Perdew–Burke–Ernzerhof functional (PAW-PBE) and a cutoff energy of 400 eV are performed. The multilayer $MoS_2$ structures are adopted by the experimental lattice of bulk 2H-$MoS_2$, and the structures are fully relaxed until the Hellmann-Feynman forces become less than 0.01 eV/Å, by using the van der Waals optB88 functional[3] to obtain an accurate description of the dispersion force and the interlayer distance. The first Brillouin zone is sampled with a $15\times15\times1$ Monkhorst-Pack grid and a vacuum region with thickness greater than 15 Å is adopted.

The optical properties are calculated based on independent-particle approximation. The involved unoccupied band number is about 10 times of that of valence bands to achieve the converged dielectric function. The accuracy of the k-point sampling is tested for using $15\times15$ and $18\times18$ grid for 1L-3L $MoS_2$, and the calculated dielectric spectra is shown in Fig. S7. While there are some discrepancies in the low-energy part between these two k-points sets, good convergence is observed for the "C" peak at around 2.75 eV, consistent with previous work[4]. Similarly, for the SOC effect (Fig. S8), the imaginary part of the dielectric constant for $MoS_2$ layers from 1L to 5L shows little effect from the SOC.

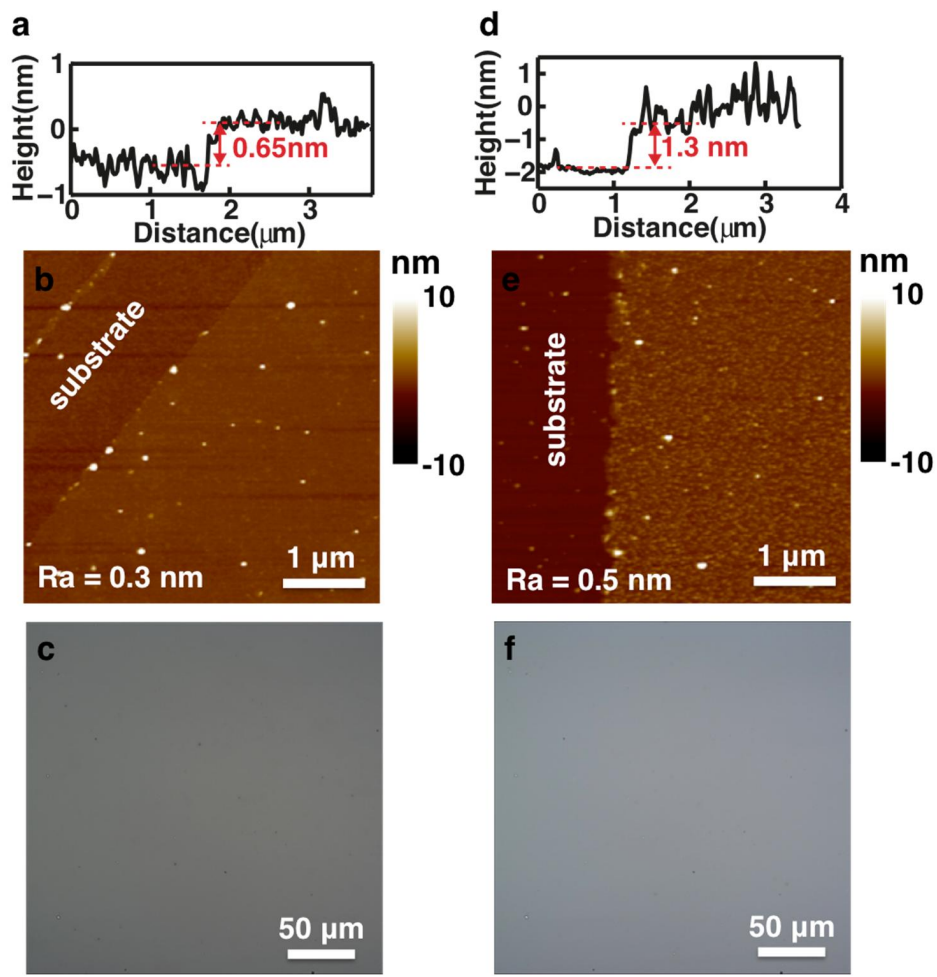

**Figure S1. Characterization of monolayer and bilayer MoS2 films. (a-c)** Typical height profile extracted from AFM measurements, AFM image, and optical image of monolayer $MoS_2$ films. The surface roughness is measured as 0.3 nm as shown in the AFM image. **(d-f)** Typical height profile extracted from AFM measurements, AFM image, and optical image of bilayer $MoS_2$ films. The surface roughness is measured as 0.5 nm as shown in the AFM image.

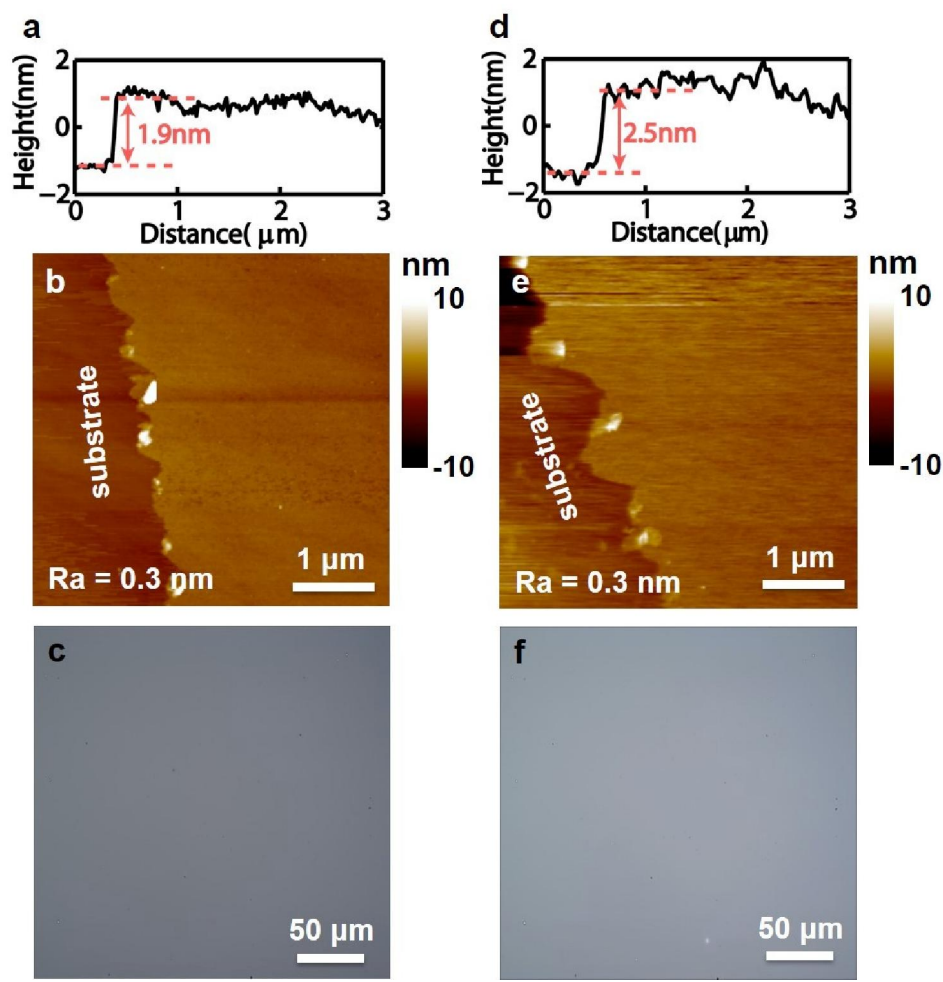

**Figure S2. Characterization of three-layer (3L) and four-layer (4L) MoS$_2$ films. (a-c)** Typical height profile extracted from AFM measurements, AFM image, and optical image of 3L MoS$_2$ films. The surface roughness is measured as 0.3 nm as shown in the AFM image. **(d-f)** Typical height profile extracted from AFM measurements, AFM image, and optical image of 4L MoS$_2$ films. The surface roughness is measured as 0.3 nm as shown in the AFM image.

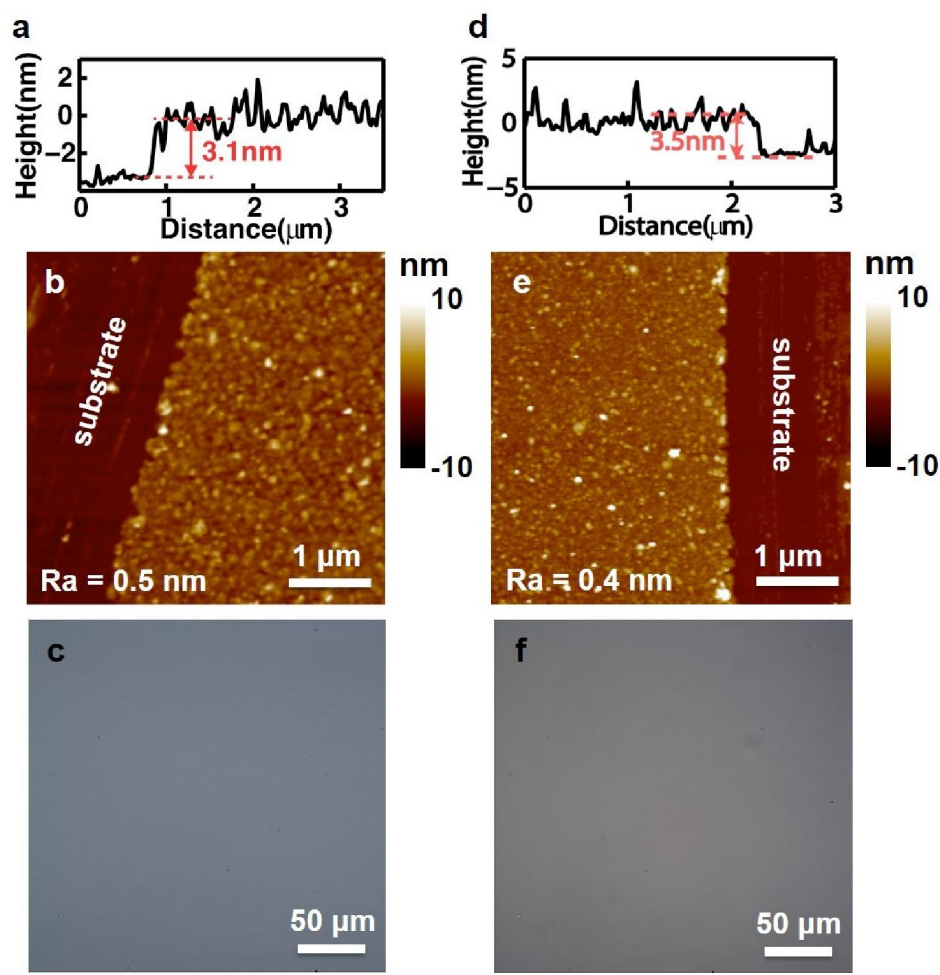

**Figure S3. Characterization of five-layer (5L) and six-layer (6L) MoS₂ films. (a-c)** Typical height profile extracted from AFM measurements, AFM image, and optical image of 5L MoS₂ films. The surface roughness is measured as 0.5 nm as shown in the AFM image. **(d-f)** Typical height profile extracted from AFM measurements, AFM image, and optical image of 6L MoS₂ films. The surface roughness is measured as 0.4 nm as shown in the AFM image.

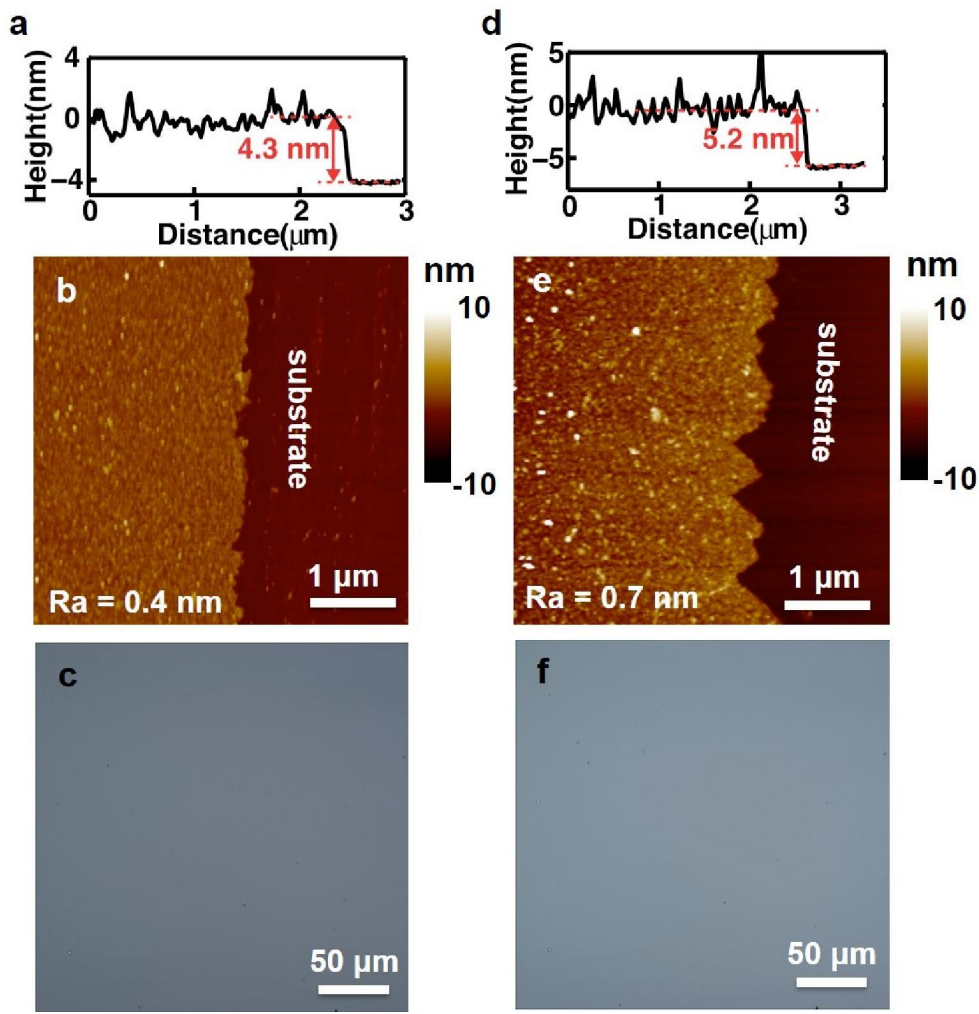

**Figure S4. Characterization of seven-layer (7L) and eight-layer (8L) MoS$_2$ films. (a-c)** Typical height profile extracted from AFM measurements, AFM image, and optical image of 7L MoS$_2$ films. The surface roughness is measured as 0.4 nm as shown in the AFM image. **(d-f)** Typical height profile extracted from AFM measurements, AFM image, and optical image of 8L MoS$_2$ films. The surface roughness is measured as 0.9 nm as shown in the AFM image

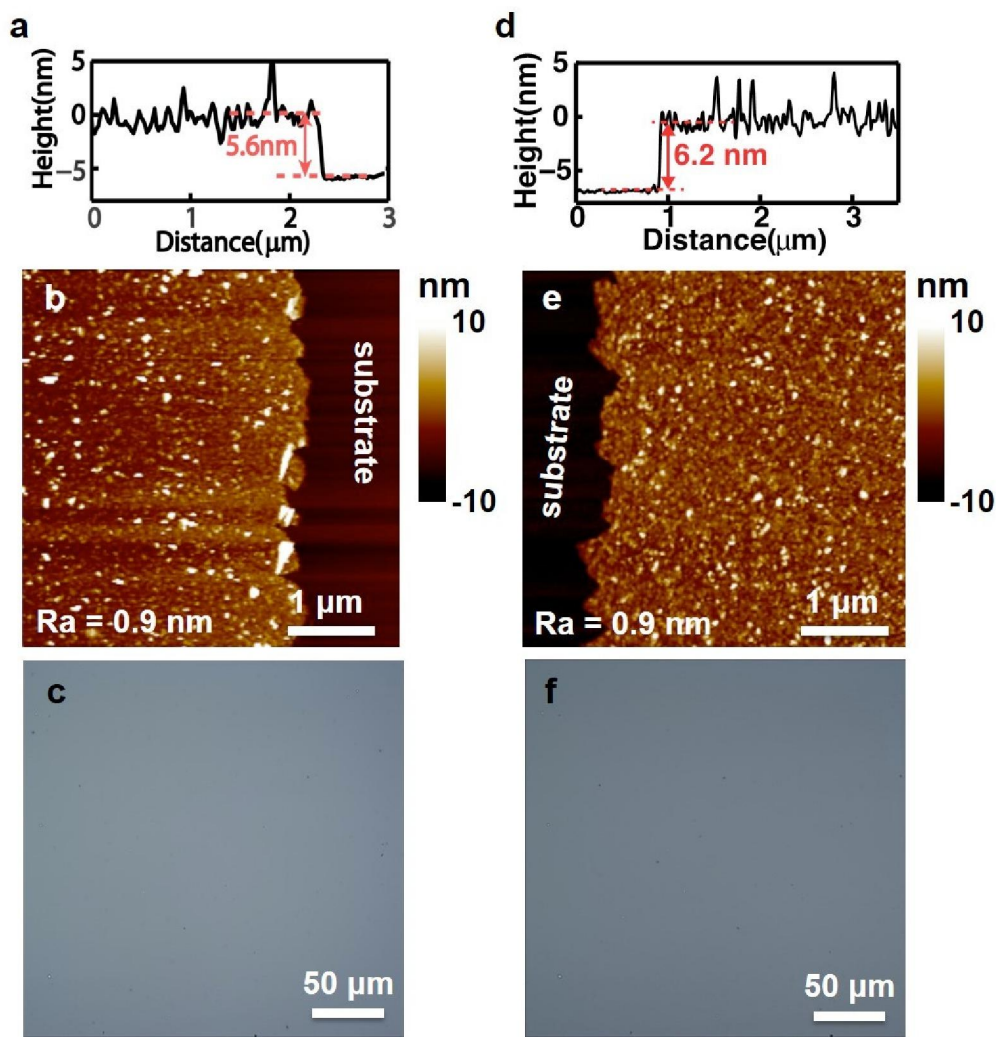

**Figure S5. Characterization of nine-layer (9L) and ten-layer (10L) MoS$_2$ films. (a-c)** Typical height profile extracted from AFM measurements, AFM image, and optical image of 9L MoS$_2$ films. The surface roughness is measured as 0.9 nm as shown in the AFM image. **(d-f)** Typical height profile extracted from AFM measurements, AFM image, and optical image of 10L MoS$_2$ films. The surface roughness is measured as 0.9 nm as shown in the AFM image

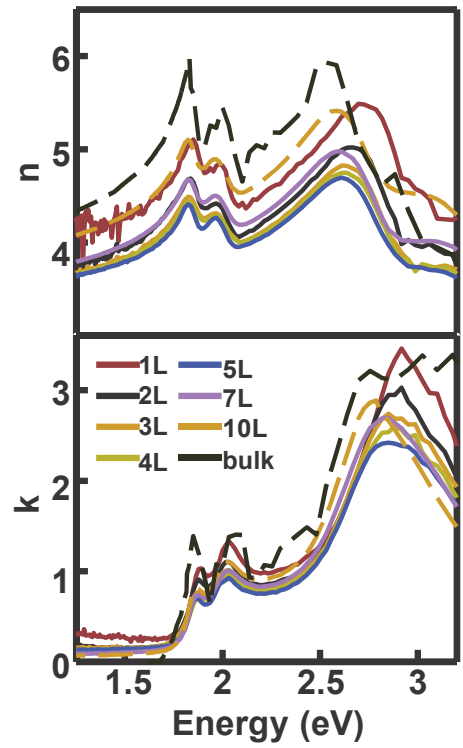

**Figure S6. Refractive index of 2D MoS$_2$.** Real (upper panel) and imaginary (lower panel) parts of the dielectric function of 2D MoS$_2$ vs. layer number. Also given is the refractive index of bulk MoS$_2$.

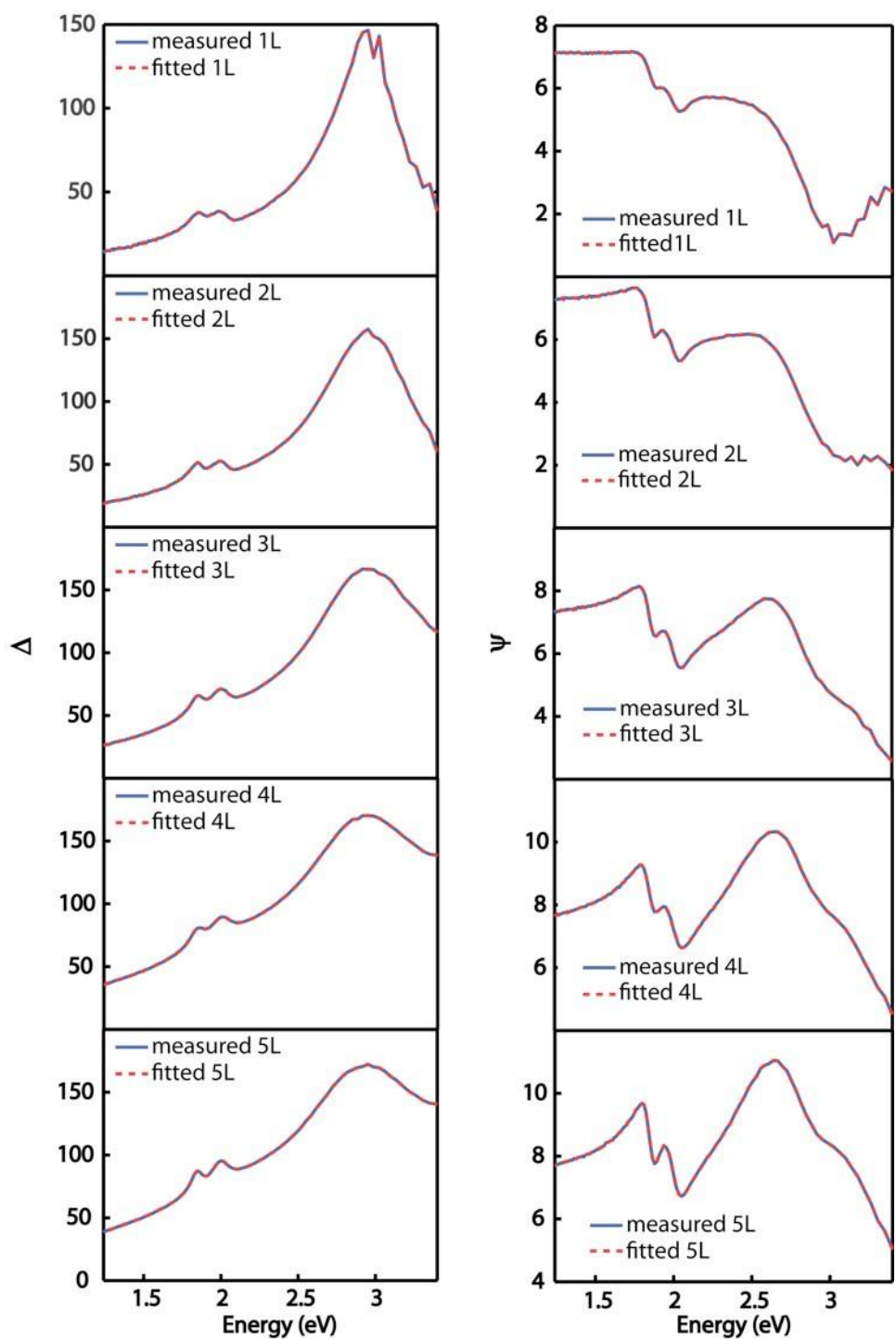

**Figure S7. Fitting for the results obtained from ellipsometry measurements.** (a) Fitted and measured phase $\Delta$ from monolayer to 5-layer MoS2 films. (b) Fitted and measured amplitude $\psi$ for monolayer to 5-layer MoS2 films.

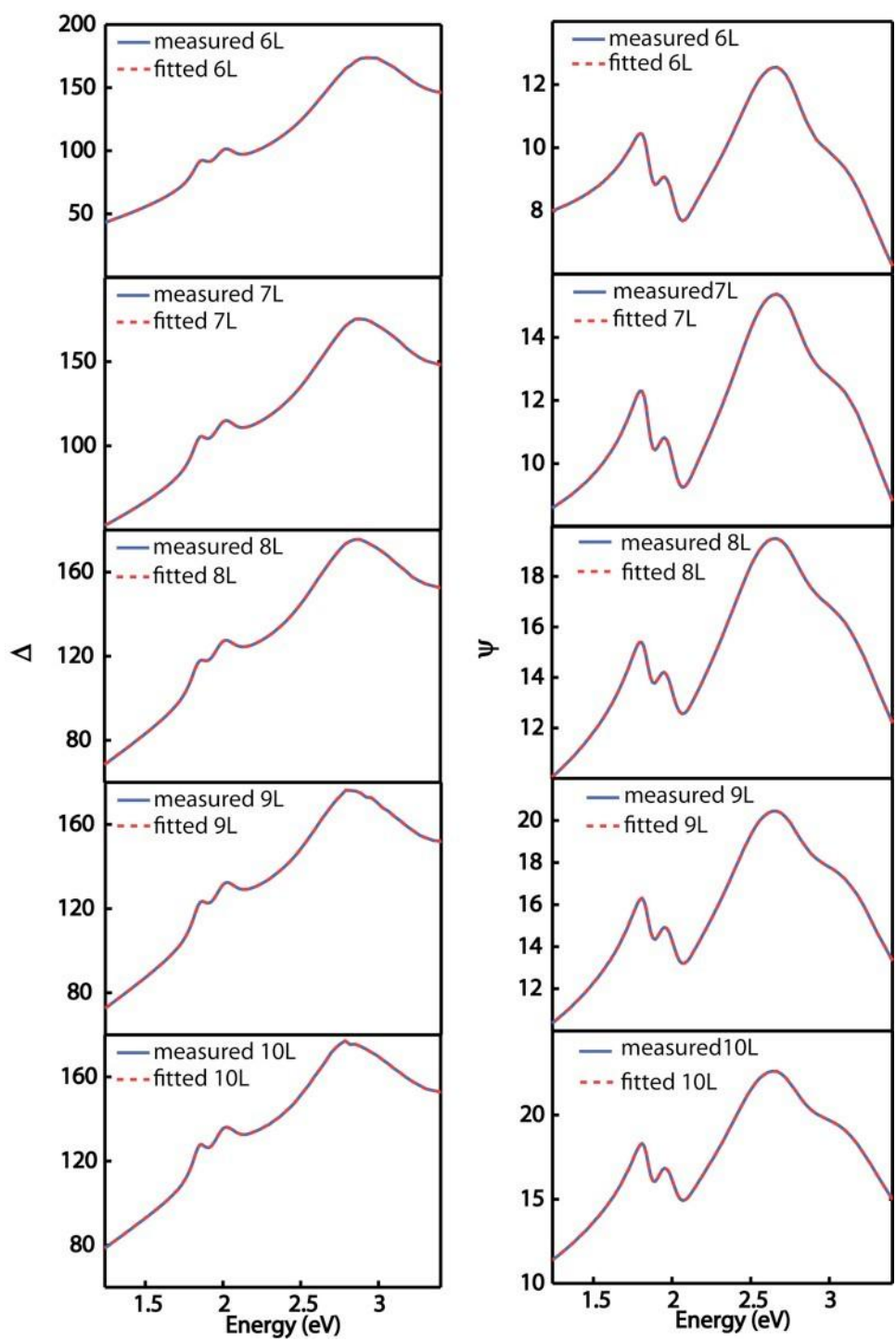

**Figure S8. Fitting for the results obtained from ellipsometry measurements.** (a) Fitted and measured phase $\Delta$ from 6-layer to 10-layer MoS2 films. (b) Fitted and measured amplitude $\psi$ for 6-layer to 10-layer MoS2 films.

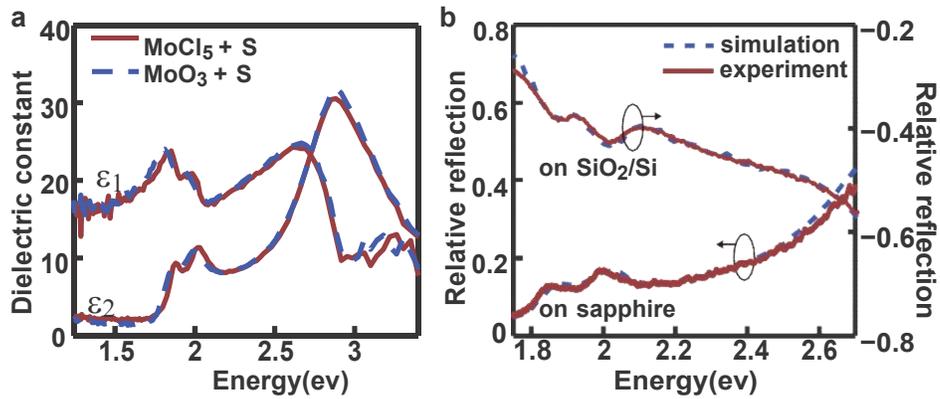

**Figure S9. Independence of the measured dielectric function from synthetic processes and substrates.** (a) Measured dielectric constants of as-grown monolayer $MoS_2$ made by two different CVD processes, one using $MoCl_5$ and S as the precursors and the other using $MoO_3$ and S as the precursors. The growth substrate in both growths is sapphire. (b) Measured and simulated reflection spectra of monolayer $MoS_2$ on different substrates, sapphire and silicon with 80 nm thick thermal oxide. The monolayer $MoS_2$ involved is grown on sapphire substrates and then transferred to $SiO_2$/Si substrates. The simulation uses the optical constant measured with the as-grown $MoS_2$ on sapphire substrates.

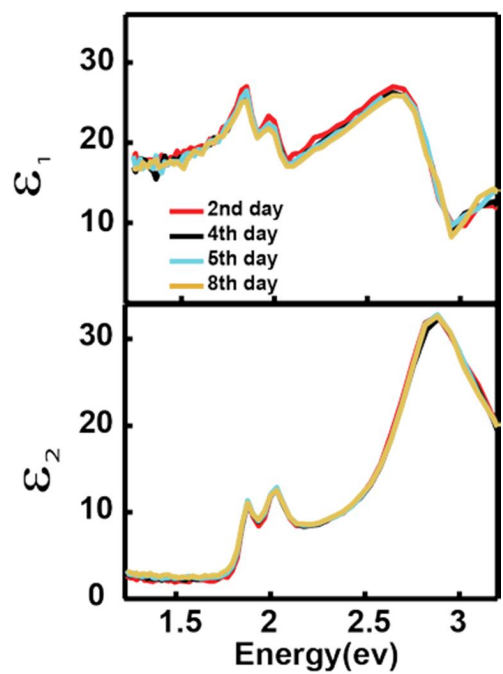

**Figure S10. Stability of the dielectric function.** Real (upper panel) and imaginary (lower panel) parts of the dielectric function of monolayer MoS$_2$ measured at different times after synthesis.

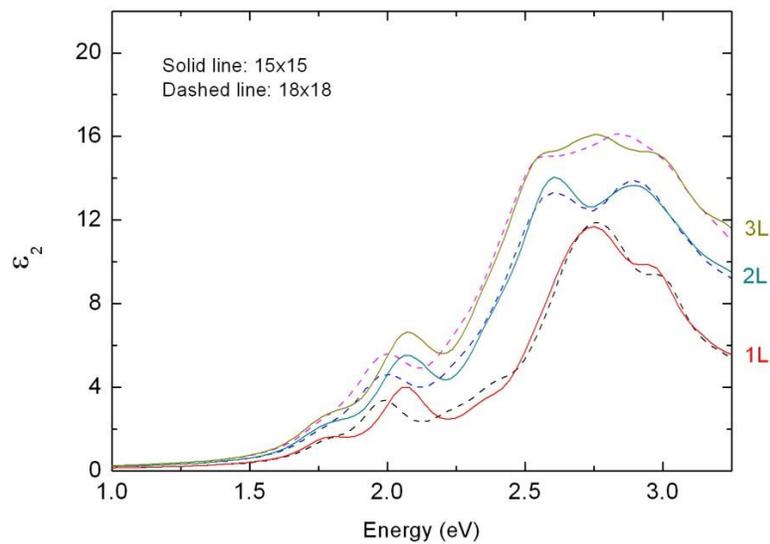

**Figure S11.** Imaginary part of the dielectric constant for MoS$_2$ layers from 1L to 3L calculated different k-point sampling.

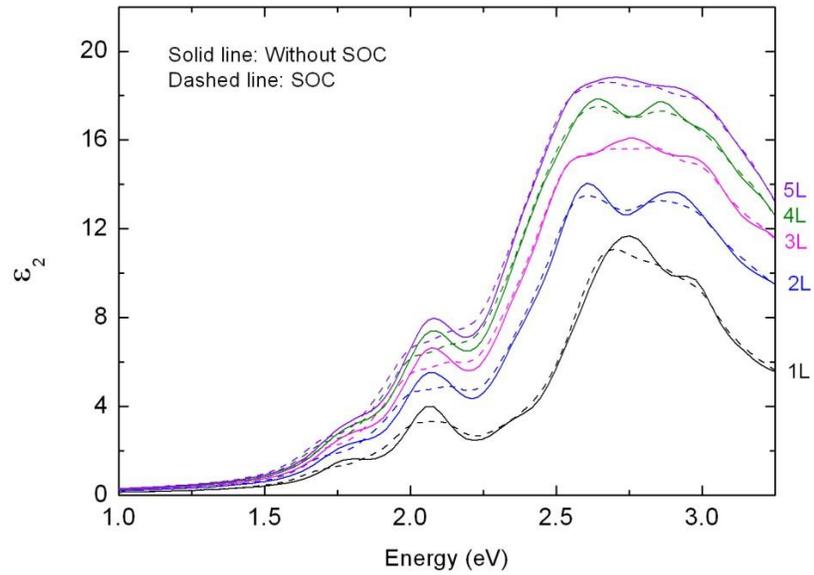

**Figure S12.** Imaginary part of the dielectric constant for $MoS_2$ layers from 1L to 5L calculated with spin-orbital coupling (dashed lines) and without spin-orbital coupling (solid lines).

## Table S1. Tabulated refractive index of MoS$_2$ films

| wavelength(nm) | 1L n | 1L k | 2L n | 2L k | 3L n | 3L k |
|---|---|---|---|---|---|---|
| 345 | 3.035 | 2.1095 | 3.1215 | 1.48 | 3.1584 | 1.4242 |
| 350 | 3.3642 | 1.711 | 3.1431 | 1.6304 | 3.1799 | 1.3707 |
| 355 | 3.2948 | 1.7682 | 3.3188 | 1.5147 | 3.1516 | 1.4211 |
| 360 | 3.0747 | 2.2111 | 3.4153 | 1.4939 | 3.3381 | 1.3626 |
| 365 | 3.4136 | 1.9955 | 3.482 | 1.4957 | 3.3745 | 1.4535 |
| 370 | 4.0257 | 1.8186 | 3.5141 | 1.636 | 3.4953 | 1.5307 |
| 375 | 3.9504 | 1.964 | 3.6118 | 1.6953 | 3.5514 | 1.6638 |
| 380 | 4.2656 | 2.1141 | 3.7339 | 1.8568 | 3.6016 | 1.8068 |
| 385 | 4.2966 | 2.2502 | 3.8891 | 1.9414 | 3.6983 | 1.8775 |
| 390 | 4.241 | 2.5232 | 3.8747 | 2.128 | 3.7274 | 1.9909 |
| 395 | 4.2479 | 2.7103 | 3.8809 | 2.3194 | 3.7528 | 2.1365 |
| 400 | 4.2558 | 2.9601 | 3.9263 | 2.4171 | 3.7368 | 2.2653 |
| 405 | 4.4645 | 2.9673 | 3.987 | 2.4962 | 3.721 | 2.4118 |
| 410 | 4.538 | 3.1094 | 3.9502 | 2.6409 | 3.7648 | 2.4578 |
| 415 | 4.5785 | 3.2506 | 3.9717 | 2.7826 | 3.7127 | 2.6254 |
| 420 | 4.6688 | 3.3453 | 4.0257 | 2.8704 | 3.816 | 2.6231 |
| 425 | 4.6942 | 3.461 | 3.9825 | 3.0278 | 3.889 | 2.6862 |
| 430 | 4.9523 | 3.334 | 4.2523 | 2.9482 | 4.0043 | 2.6842 |
| 435 | 5.1659 | 3.1845 | 4.3822 | 2.9453 | 4.0705 | 2.7412 |
| 440 | 5.3321 | 2.9496 | 4.5641 | 2.8467 | 4.2579 | 2.6839 |
| 445 | 5.4295 | 2.7198 | 4.724 | 2.6996 | 4.4327 | 2.5853 |
| 450 | 5.471 | 2.5039 | 4.8808 | 2.5191 | 4.5802 | 2.4602 |
| 455 | 5.4828 | 2.3085 | 4.9702 | 2.3155 | 4.6858 | 2.2951 |
| 460 | 5.4897 | 2.1087 | 5.0077 | 2.128 | 4.7639 | 2.1338 |
| 465 | 5.4255 | 1.9334 | 5.0223 | 1.9605 | 4.8122 | 1.9741 |
| 470 | 5.3851 | 1.7774 | 5.0186 | 1.7949 | 4.8247 | 1.802 |
| 475 | 5.3511 | 1.6549 | 4.9916 | 1.648 | 4.8261 | 1.6554 |
| 480 | 5.2685 | 1.5181 | 4.9396 | 1.509 | 4.7934 | 1.5116 |
| 485 | 5.1999 | 1.4334 | 4.9015 | 1.3921 | 4.7569 | 1.3899 |
| 490 | 5.1296 | 1.3522 | 4.8236 | 1.2948 | 4.7065 | 1.2803 |
| 495 | 5.0832 | 1.2878 | 4.7654 | 1.2128 | 4.6505 | 1.1969 |
| 500 | 5.0093 | 1.2361 | 4.7054 | 1.1513 | 4.5976 | 1.1257 |
| 505 | 4.9526 | 1.1885 | 4.6687 | 1.0974 | 4.5475 | 1.0737 |
| 510 | 4.918 | 1.1623 | 4.6115 | 1.0532 | 4.4963 | 1.0202 |
| 515 | 4.8614 | 1.1215 | 4.5572 | 1.0157 | 4.4548 | 0.97899 |
| 520 | 4.8404 | 1.0768 | 4.511 | 0.98054 | 4.4037 | 0.93817 |
| 525 | 4.7757 | 1.0698 | 4.4642 | 0.94246 | 4.354 | 0.91249 |
| 530 | 4.7458 | 1.0402 | 4.4245 | 0.92324 | 4.3213 | 0.89069 |
| 535 | 4.6875 | 1.0243 | 4.3847 | 0.91436 | 4.2826 | 0.87048 |
| 540 | 4.6276 | 1.0169 | 4.3481 | 0.8949 | 4.2396 | 0.85457 |
| 545 | 4.6184 | 1.0045 | 4.3139 | 0.87514 | 4.2103 | 0.84473 |
| 550 | 4.5974 | 0.98786 | 4.2807 | 0.86571 | 4.1887 | 0.83567 |
| 555 | 4.5692 | 0.97655 | 4.2529 | 0.8587 | 4.151 | 0.82515 |
| 560 | 4.5375 | 0.98979 | 4.2331 | 0.84906 | 4.1273 | 0.81697 |
| 565 | 4.4953 | 0.97569 | 4.1915 | 0.84723 | 4.1036 | 0.81305 |
| 570 | 4.4593 | 0.98802 | 4.1599 | 0.85819 | 4.0785 | 0.8257 |
| 575 | 4.4652 | 0.99232 | 4.1311 | 0.86358 | 4.0557 | 0.83162 |
| 580 | 4.4046 | 1.0283 | 4.1233 | 0.86556 | 4.0351 | 0.84601 |
| 585 | 4.3665 | 1.0652 | 4.0993 | 0.90042 | 4.0186 | 0.85957 |
| 590 | 4.4281 | 1.1133 | 4.0773 | 0.92505 | 4.0076 | 0.88298 |
| 595 | 4.3854 | 1.1736 | 4.0568 | 0.98182 | 4.007 | 0.91261 |
| 600 | 4.4185 | 1.2453 | 4.0802 | 1.0272 | 4.0194 | 0.95695 |
| 605 | 4.4832 | 1.293 | 4.1333 | 1.0777 | 4.0487 | 0.98958 |
| 610 | 4.5949 | 1.3319 | 4.1995 | 1.1032 | 4.1102 | 1.0138 |
| 615 | 4.7645 | 1.2887 | 4.2755 | 1.1024 | 4.1728 | 1.0085 |
| 620 | 4.823 | 1.2336 | 4.3632 | 1.0654 | 4.238 | 0.97982 |
| 625 | 4.8473 | 1.1104 | 4.405 | 0.99226 | 4.2943 | 0.9388 |
| 630 | 4.8233 | 1.0397 | 4.4202 | 0.93166 | 4.3156 | 0.8626 |
| 635 | 4.7608 | 0.9785 | 4.4086 | 0.87284 | 4.3128 | 0.80623 |
| 640 | 4.7993 | 0.9588 | 4.3911 | 0.84114 | 4.295 | 0.76433 |
| 645 | 4.757 | 0.94393 | 4.3689 | 0.82963 | 4.2724 | 0.7384 |
| 650 | 4.7445 | 0.97564 | 4.3675 | 0.84836 | 4.2619 | 0.74509 |
| 655 | 4.8177 | 1.033 | 4.3777 | 0.87778 | 4.2712 | 0.76408 |
| 660 | 4.9334 | 1.0246 | 4.448 | 0.90742 | 4.3095 | 0.77144 |
| 665 | 5.0486 | 0.94987 | 4.5632 | 0.87405 | 4.373 | 0.75421 |
| 670 | 5.1062 | 0.80879 | 4.6477 | 0.78976 | 4.4384 | 0.70404 |
| 675 | 5.0349 | 0.67168 | 4.6851 | 0.66598 | 4.4861 | 0.62352 |
| 680 | 5.0504 | 0.53835 | 4.6786 | 0.54976 | 4.4922 | 0.52238 |
| 685 | 4.9719 | 0.45368 | 4.6382 | 0.47019 | 4.4656 | 0.42973 |
| 690 | 4.9161 | 0.39993 | 4.5794 | 0.39449 | 4.4248 | 0.365 |
| 695 | 4.8819 | 0.34722 | 4.5294 | 0.34297 | 4.3719 | 0.31735 |
| 700 | 4.7792 | 0.31574 | 4.4932 | 0.30126 | 4.3385 | 0.28054 |
| 705 | 4.7406 | 0.3121 | 4.4461 | 0.27104 | 4.2891 | 0.2527 |
| 710 | 4.7011 | 0.3007 | 4.4192 | 0.24924 | 4.2509 | 0.23223 |
| 715 | 4.6449 | 0.28574 | 4.3796 | 0.21906 | 4.2112 | 0.21784 |
| 720 | 4.6281 | 0.2684 | 4.3454 | 0.20235 | 4.1777 | 0.20581 |
| 725 | 4.6008 | 0.25389 | 4.3018 | 0.18963 | 4.149 | 0.19284 |
| 730 | 4.5917 | 0.26534 | 4.2665 | 0.18329 | 4.1269 | 0.18748 |
| 735 | 4.564 | 0.25114 | 4.2493 | 0.17875 | 4.1062 | 0.18071 |
| 740 | 4.4731 | 0.25553 | 4.2084 | 0.17008 | 4.0695 | 0.1786 |
| 745 | 4.4781 | 0.23703 | 4.1905 | 0.16271 | 4.0515 | 0.17313 |
| 750 | 4.4968 | 0.26643 | 4.1663 | 0.1677 | 4.0331 | 0.16169 |
| 755 | 4.4555 | 0.24217 | 4.1474 | 0.16509 | 4.0284 | 0.16304 |
| 760 | 4.4546 | 0.25444 | 4.1122 | 0.15967 | 4.0049 | 0.1699 |
| 765 | 4.4104 | 0.25882 | 4.108 | 0.15712 | 3.9894 | 0.1553 |
| 770 | 4.3872 | 0.25604 | 4.0846 | 0.15634 | 3.9607 | 0.1585 |
| 775 | 4.3246 | 0.2661 | 4.0686 | 0.16126 | 3.9767 | 0.16305 |
| 780 | 4.3012 | 0.22656 | 4.0303 | 0.15816 | 3.9409 | 0.1541 |
| 785 | 4.3975 | 0.27913 | 4.0053 | 0.16128 | 3.9261 | 0.15155 |
| 790 | 4.3033 | 0.23178 | 4.0308 | 0.15668 | 3.8957 | 0.1591 |
| 795 | 4.3539 | 0.25401 | 3.9956 | 0.15633 | 3.882 | 0.16254 |
| 800 | 4.3375 | 0.2661 | 3.9946 | 0.1528 | 3.8753 | 0.14879 |
| 805 | 4.2829 | 0.27118 | 3.9805 | 0.15312 | 3.8912 | 0.15733 |
| 810 | 4.2618 | 0.21999 | 3.963 | 0.14757 | 3.871 | 0.14949 |
| 815 | 4.4143 | 0.24665 | 3.9551 | 0.15666 | 3.8526 | 0.15579 |
| 820 | 4.262 | 0.25912 | 3.9322 | 0.16006 | 3.8459 | 0.16055 |
| 825 | 4.2599 | 0.25859 | 3.9248 | 0.15353 | 3.837 | 0.15222 |
| 830 | 4.2536 | 0.28422 | 3.9241 | 0.14709 | 3.843 | 0.15066 |
| 835 | 4.3063 | 0.27486 | 3.9432 | 0.1558 | 3.8332 | 0.14969 |
| 840 | 4.2071 | 0.26591 | 3.9436 | 0.15872 | 3.8175 | 0.15397 |
| 845 | 4.0152 | 0.30428 | 3.9045 | 0.15062 | 3.8431 | 0.14614 |
| 850 | 4.384 | 0.28634 | 3.9846 | 0.16584 | 3.8452 | 0.17482 |
| 855 | 4.2731 | 0.27395 | 3.9277 | 0.16079 | 3.7805 | 0.1484 |
| 860 | 4.2313 | 0.29451 | 3.9019 | 0.16303 | 3.8043 | 0.16181 |
| 865 | 4.0518 | 0.25802 | 3.8836 | 0.16324 | 3.8042 | 0.16507 |
| 870 | 4.2107 | 0.29103 | 3.8418 | 0.159 | 3.7678 | 0.15209 |
| 875 | 4.1901 | 0.28231 | 3.8546 | 0.15625 | 3.7828 | 0.14107 |
| 880 | 4.2146 | 0.2688 | 3.8225 | 0.15519 | 3.7643 | 0.16109 |
| 885 | 4.1368 | 0.26439 | 3.8191 | 0.15187 | 3.7538 | 0.14589 |
| 890 | 4.2505 | 0.27241 | 3.8828 | 0.15811 | 3.7618 | 0.14685 |
| 895 | 4.1246 | 0.275 | 3.823 | 0.15444 | 3.7427 | 0.15058 |
| 900 | 4.2324 | 0.27653 | 3.8394 | 0.16086 | 3.7517 | 0.15551 |
| 905 | 4.1399 | 0.29945 | 3.8437 | 0.15368 | 3.7459 | 0.1452 |
| 910 | 4.0882 | 0.27867 | 3.8041 | 0.14879 | 3.7418 | 0.16137 |
| 915 | 4.204 | 0.27915 | 3.8372 | 0.16322 | 3.7354 | 0.15593 |
| 920 | 4.2151 | 0.29217 | 3.7608 | 0.1558 | 3.7179 | 0.14731 |
| 925 | 4.1818 | 0.31775 | 3.8349 | 0.16625 | 3.7208 | 0.15206 |
| 930 | 3.9989 | 0.27231 | 3.8169 | 0.17591 | 3.727 | 0.14932 |
| 935 | 4.2799 | 0.28077 | 3.8124 | 0.16884 | 3.6963 | 0.15023 |
| 940 | 4.2274 | 0.30284 | 3.7828 | 0.16941 | 3.6996 | 0.16622 |
| 945 | 4.3511 | 0.29421 | 3.7913 | 0.16523 | 3.7445 | 0.15839 |
| 950 | 4.2508 | 0.29206 | 3.7844 | 0.17303 | 3.6923 | 0.15044 |
| 955 | 4.3307 | 0.29048 | 3.8152 | 0.16167 | 3.7089 | 0.15293 |
| 960 | 4.1119 | 0.36195 | 3.7906 | 0.17346 | 3.6594 | 0.15831 |
| 965 | 3.6847 | 0.30302 | 3.7358 | 0.16777 | 3.6834 | 0.16439 |
| 970 | 4.2891 | 0.30635 | 3.7615 | 0.17787 | 3.6761 | 0.15848 |
| 975 | 4.2642 | 0.32016 | 3.7736 | 0.18801 | 3.7132 | 0.16278 |

| wavelength(nm) | | | | | | |
|---|---|---|---|---|---|---|
| 980 | 4.0643 | 0.30154 | 3.7598 | 0.16946 | 3.6674 | 0.15559 |
| 985 | 4.0738 | 0.30023 | 3.7561 | 0.18424 | 3.6723 | 0.15389 |
| 990 | 4.26 | 0.31443 | 3.8044 | 0.17347 | 3.68 | 0.15773 |
| 995 | 4.1482 | 0.32433 | 3.7509 | 0.18282 | 3.6638 | 0.15621 |
| 1000 | 4.1655 | 0.32855 | 3.7309 | 0.1854 | 3.6428 | 0.15677 |

| | 4L | | 5L | | 6L | |
|---|---|---|---|---|---|---|
| wavelength(nm) | n | k | n | k | n | k |
| 345 | 2.5614 | 1.8298 | 3.0358 | 1.3321 | 3.0844 | 1.3501 |
| 350 | 3.194 | 1.3347 | 3.0551 | 1.3281 | 3.1263 | 1.3278 |
| 355 | 3.125 | 1.4035 | 3.1296 | 1.3226 | 3.1888 | 1.3263 |
| 360 | 3.2429 | 1.4139 | 3.2271 | 1.321 | 3.2635 | 1.3337 |
| 365 | 3.2509 | 1.5094 | 3.2787 | 1.3537 | 3.3504 | 1.3692 |
| 370 | 3.4527 | 1.4604 | 3.3782 | 1.4042 | 3.4356 | 1.4218 |
| 375 | 3.5296 | 1.5183 | 3.4467 | 1.4665 | 3.5215 | 1.4918 |
| 380 | 3.5911 | 1.6489 | 3.5307 | 1.572 | 3.598 | 1.5798 |
| 385 | 3.6367 | 1.7441 | 3.5988 | 1.6661 | 3.6692 | 1.6879 |
| 390 | 3.694 | 1.8858 | 3.6456 | 1.7875 | 3.7171 | 1.8181 |
| 395 | 3.7279 | 2.0512 | 3.6812 | 1.9205 | 3.7502 | 1.9475 |
| 400 | 3.7228 | 2.1625 | 3.7 | 2.0316 | 3.7609 | 2.0817 |
| 405 | 3.7459 | 2.205 | 3.7055 | 2.1543 | 3.7696 | 2.1851 |
| 410 | 3.7663 | 2.3166 | 3.7282 | 2.2037 | 3.7707 | 2.2849 |
| 415 | 3.689 | 2.5041 | 3.716 | 2.311 | 3.7685 | 2.3836 |
| 420 | 3.8204 | 2.4765 | 3.7435 | 2.373 | 3.8037 | 2.4242 |
| 425 | 3.825 | 2.5621 | 3.8057 | 2.3871 | 3.8434 | 2.4684 |
| 430 | 3.9221 | 2.6016 | 3.8774 | 2.4095 | 3.9145 | 2.5102 |
| 435 | 4.0918 | 2.5361 | 3.9761 | 2.4199 | 4.025 | 2.5045 |
| 440 | 4.2156 | 2.5281 | 4.0915 | 2.4114 | 4.1697 | 2.4642 |
| 445 | 4.3628 | 2.4465 | 4.2231 | 2.3754 | 4.3019 | 2.4399 |
| 450 | 4.4941 | 2.3408 | 4.363 | 2.3038 | 4.4553 | 2.3357 |
| 455 | 4.6071 | 2.1983 | 4.4893 | 2.1779 | 4.5806 | 2.2191 |
| 460 | 4.6802 | 2.0611 | 4.584 | 2.0393 | 4.678 | 2.0717 |
| 465 | 4.7282 | 1.8988 | 4.6426 | 1.8974 | 4.7455 | 1.9145 |
| 470 | 4.749 | 1.7469 | 4.6799 | 1.7471 | 4.7787 | 1.762 |
| 475 | 4.7481 | 1.6091 | 4.696 | 1.5934 | 4.789 | 1.619 |
| 480 | 4.7362 | 1.4717 | 4.6796 | 1.4624 | 4.7778 | 1.4747 |
| 485 | 4.7031 | 1.3579 | 4.6507 | 1.3532 | 4.7532 | 1.3492 |
| 490 | 4.6591 | 1.2501 | 4.6138 | 1.2294 | 4.7125 | 1.2375 |
| 495 | 4.6139 | 1.1679 | 4.57 | 1.156 | 4.6664 | 1.145 |
| 500 | 4.5608 | 1.0933 | 4.5203 | 1.0741 | 4.614 | 1.068 |
| 505 | 4.5105 | 1.0365 | 4.4705 | 1.0094 | 4.562 | 1.0016 |
| 510 | 4.4681 | 0.98793 | 4.4263 | 0.95779 | 4.5105 | 0.94978 |
| 515 | 4.4167 | 0.94254 | 4.3782 | 0.91445 | 4.4636 | 0.90615 |
| 520 | 4.3717 | 0.91075 | 4.337 | 0.87585 | 4.416 | 0.86924 |
| 525 | 4.3385 | 0.87681 | 4.2945 | 0.85184 | 4.3725 | 0.83965 |
| 530 | 4.2953 | 0.85398 | 4.2561 | 0.82294 | 4.3307 | 0.81397 |
| 535 | 4.2566 | 0.83392 | 4.2181 | 0.80078 | 4.2934 | 0.79583 |
| 540 | 4.2195 | 0.82141 | 4.1836 | 0.78992 | 4.2575 | 0.77728 |
| 545 | 4.196 | 0.803 | 4.1526 | 0.7659 | 4.2234 | 0.76694 |
| 550 | 4.1629 | 0.79384 | 4.1246 | 0.76979 | 4.1946 | 0.75332 |
| 555 | 4.1337 | 0.79206 | 4.0911 | 0.75762 | 4.165 | 0.74949 |
| 560 | 4.1085 | 0.78915 | 4.0697 | 0.75156 | 4.1388 | 0.74522 |
| 565 | 4.0855 | 0.78346 | 4.0452 | 0.75376 | 4.1148 | 0.74631 |
| 570 | 4.0621 | 0.78814 | 4.0281 | 0.75084 | 4.0906 | 0.74932 |
| 575 | 4.0469 | 0.79079 | 4.0067 | 0.76026 | 4.0728 | 0.75755 |
| 580 | 4.0238 | 0.80428 | 3.9861 | 0.76684 | 4.0539 | 0.7701 |
| 585 | 4.0138 | 0.81894 | 3.9703 | 0.77828 | 4.0419 | 0.78739 |
| 590 | 4.0034 | 0.84196 | 3.9569 | 0.79617 | 4.0334 | 0.81297 |
| 595 | 4.0026 | 0.87146 | 3.9538 | 0.8258 | 4.0393 | 0.8412 |
| 600 | 4.0203 | 0.91388 | 3.9554 | 0.8607 | 4.0595 | 0.87293 |
| 605 | 4.0568 | 0.94228 | 3.9835 | 0.90045 | 4.0972 | 0.89817 |
| 610 | 4.1131 | 0.95913 | 4.0276 | 0.92813 | 4.1509 | 0.9087 |
| 615 | 4.1722 | 0.94849 | 4.1008 | 0.93788 | 4.2138 | 0.89697 |
| 620 | 4.2366 | 0.91014 | 4.1759 | 0.91607 | 4.2678 | 0.85902 |
| 625 | 4.272 | 0.85979 | 4.2334 | 0.85543 | 4.3055 | 0.80191 |
| 630 | 4.287 | 0.79769 | 4.2669 | 0.78154 | 4.3149 | 0.74104 |
| 635 | 4.2841 | 0.74063 | 4.2544 | 0.70846 | 4.3088 | 0.68818 |
| 640 | 4.2645 | 0.70184 | 4.2269 | 0.65469 | 4.2849 | 0.65194 |
| 645 | 4.2443 | 0.69144 | 4.1877 | 0.63311 | 4.2652 | 0.638 |
| 650 | 4.2341 | 0.69161 | 4.1576 | 0.6314 | 4.2554 | 0.64115 |
| 655 | 4.2472 | 0.69864 | 4.1489 | 0.65775 | 4.2645 | 0.65279 |
| 660 | 4.2836 | 0.70641 | 4.1709 | 0.69372 | 4.3021 | 0.65914 |
| 665 | 4.3376 | 0.68971 | 4.2397 | 0.70046 | 4.3599 | 0.64246 |
| 670 | 4.4007 | 0.63422 | 4.3282 | 0.67378 | 4.4217 | 0.58933 |
| 675 | 4.431 | 0.55125 | 4.3912 | 0.58729 | 4.4573 | 0.50454 |
| 680 | 4.4362 | 0.46543 | 4.4087 | 0.47751 | 4.4554 | 0.41062 |
| 685 | 4.4055 | 0.37922 | 4.3839 | 0.37166 | 4.4243 | 0.32837 |
| 690 | 4.3647 | 0.32056 | 4.3318 | 0.30779 | 4.378 | 0.26886 |
| 695 | 4.3169 | 0.27772 | 4.2695 | 0.25896 | 4.3291 | 0.22579 |
| 700 | 4.2732 | 0.24239 | 4.2228 | 0.22526 | 4.2818 | 0.19555 |
| 705 | 4.2329 | 0.21952 | 4.1818 | 0.20342 | 4.2399 | 0.17523 |
| 710 | 4.197 | 0.20274 | 4.1423 | 0.19703 | 4.2011 | 0.16097 |
| 715 | 4.1634 | 0.19636 | 4.1053 | 0.18286 | 4.1676 | 0.15099 |
| 720 | 4.1348 | 0.18719 | 4.0859 | 0.17904 | 4.1392 | 0.14232 |
| 725 | 4.1079 | 0.17553 | 4.0524 | 0.17421 | 4.1128 | 0.13598 |
| 730 | 4.0839 | 0.17365 | 4.0299 | 0.16416 | 4.0894 | 0.13149 |
| 735 | 4.0672 | 0.16776 | 4.0156 | 0.16941 | 4.067 | 0.12855 |
| 740 | 4.0394 | 0.15977 | 3.9852 | 0.16233 | 4.0464 | 0.1255 |
| 745 | 4.015 | 0.15476 | 3.9688 | 0.16532 | 4.0261 | 0.12153 |
| 750 | 4.0067 | 0.15455 | 3.9592 | 0.1583 | 4.0097 | 0.11965 |
| 755 | 3.979 | 0.14984 | 3.9319 | 0.15873 | 3.9921 | 0.11762 |
| 760 | 3.9687 | 0.149 | 3.926 | 0.15605 | 3.9749 | 0.11452 |
| 765 | 3.9546 | 0.14577 | 3.9118 | 0.15454 | 3.9617 | 0.11422 |
| 770 | 3.9386 | 0.14733 | 3.899 | 0.14716 | 3.9498 | 0.11075 |
| 775 | 3.9217 | 0.1442 | 3.8898 | 0.15178 | 3.9342 | 0.10927 |
| 780 | 3.9136 | 0.13708 | 3.8721 | 0.14377 | 3.9212 | 0.10985 |
| 785 | 3.9127 | 0.14229 | 3.8594 | 0.14723 | 3.9133 | 0.10929 |
| 790 | 3.8948 | 0.14163 | 3.8463 | 0.14252 | 3.9023 | 0.10914 |
| 795 | 3.8853 | 0.14457 | 3.8369 | 0.14951 | 3.8884 | 0.10814 |
| 800 | 3.8607 | 0.1339 | 3.8281 | 0.14275 | 3.8839 | 0.10661 |
| 805 | 3.8493 | 0.14862 | 3.8242 | 0.14704 | 3.8699 | 0.10802 |
| 810 | 3.8543 | 0.13833 | 3.8152 | 0.14599 | 3.8588 | 0.105 |
| 815 | 3.8419 | 0.13944 | 3.8103 | 0.14411 | 3.8526 | 0.10372 |
| 820 | 3.8366 | 0.1377 | 3.7904 | 0.14833 | 3.8446 | 0.10301 |
| 825 | 3.8218 | 0.13384 | 3.7966 | 0.13969 | 3.8375 | 0.10308 |
| 830 | 3.8188 | 0.13272 | 3.7942 | 0.14769 | 3.8305 | 0.10294 |
| 835 | 3.8049 | 0.1355 | 3.7772 | 0.14329 | 3.8229 | 0.10139 |
| 840 | 3.7957 | 0.13606 | 3.777 | 0.13432 | 3.8134 | 0.10132 |
| 845 | 3.7814 | 0.14103 | 3.7651 | 0.1462 | 3.8065 | 0.10206 |
| 850 | 3.7983 | 0.13817 | 3.7593 | 0.14139 | 3.8024 | 0.10217 |
| 855 | 3.7873 | 0.12976 | 3.7534 | 0.1437 | 3.7923 | 0.097761 |
| 860 | 3.7721 | 0.12802 | 3.7489 | 0.14498 | 3.7865 | 0.098688 |
| 865 | 3.7609 | 0.13967 | 3.7459 | 0.13894 | 3.7829 | 0.10117 |
| 870 | 3.7525 | 0.13064 | 3.7407 | 0.14129 | 3.7779 | 0.10089 |
| 875 | 3.7504 | 0.13277 | 3.733 | 0.13208 | 3.7697 | 0.098302 |
| 880 | 3.749 | 0.13245 | 3.7339 | 0.13308 | 3.7645 | 0.098093 |
| 885 | 3.7494 | 0.12771 | 3.7245 | 0.13506 | 3.7597 | 0.09857 |
| 890 | 3.7507 | 0.12503 | 3.7133 | 0.12863 | 3.7546 | 0.098106 |
| 895 | 3.7348 | 0.12769 | 3.7121 | 0.13275 | 3.7498 | 0.097763 |
| 900 | 3.7305 | 0.13342 | 3.7048 | 0.13633 | 3.7443 | 0.098424 |
| 905 | 3.7231 | 0.13129 | 3.7045 | 0.13451 | 3.7398 | 0.097871 |
| 910 | 3.7173 | 0.13197 | 3.7065 | 0.13914 | 3.7359 | 0.097249 |
| 915 | 3.7126 | 0.13251 | 3.6967 | 0.13766 | 3.7313 | 0.097509 |
| 920 | 3.7056 | 0.13923 | 3.6897 | 0.13257 | 3.727 | 0.097305 |
| 925 | 3.7127 | 0.13301 | 3.6875 | 0.13715 | 3.7229 | 0.098209 |
| 930 | 3.7068 | 0.12879 | 3.6928 | 0.13281 | 3.7178 | 0.097795 |
| 935 | 3.6978 | 0.12756 | 3.6761 | 0.1361 | 3.7143 | 0.099327 |
| 940 | 3.6934 | 0.13052 | 3.6799 | 0.12969 | 3.7109 | 0.097141 |
| 945 | 3.6914 | 0.12977 | 3.6765 | 0.12749 | 3.7078 | 0.09625 |
| 950 | 3.6886 | 0.13782 | 3.6737 | 0.12456 | 3.7051 | 0.098084 |
| 955 | 3.6894 | 0.13452 | 3.6708 | 0.13453 | 3.6999 | 0.097999 |

| wavelength(nm) | | | | | | |
|---|---|---|---|---|---|---|
| 960 | 3.6857 | 0.1339 | 3.66 | 0.12888 | 3.6974 | 0.097411 |
| 965 | 3.6795 | 0.12964 | 3.6607 | 0.13504 | 3.6908 | 0.097385 |
| 970 | 3.6736 | 0.13581 | 3.6509 | 0.13255 | 3.6874 | 0.096856 |
| 975 | 3.683 | 0.13931 | 3.6644 | 0.13727 | 3.6842 | 0.096808 |
| 980 | 3.6674 | 0.1338 | 3.6545 | 0.13241 | 3.6788 | 0.097124 |
| 985 | 3.6602 | 0.13559 | 3.6325 | 0.1357 | 3.6796 | 0.099659 |
| 990 | 3.6631 | 0.13305 | 3.6507 | 0.13053 | 3.6755 | 0.098649 |
| 995 | 3.656 | 0.13371 | 3.6474 | 0.13157 | 3.6729 | 0.099293 |
| 1000 | 3.6531 | 0.13736 | 3.6341 | 0.13545 | 3.6696 | 0.099894 |

| | 7L | | 8L | | 9L | |
|---|---|---|---|---|---|---|
| wavelength(nm) | n | k | n | k | n | k |
| 290 | 3.4972 | 1.2998 | 3.6923 | 1.1447 | 3.7718 | 1.0328 |
| 295 | 3.5023 | 1.367 | 3.7013 | 1.178 | 3.7427 | 1.1995 |
| 300 | 3.4694 | 1.394 | 3.5908 | 1.3551 | 3.6599 | 1.2548 |
| 305 | 3.4036 | 1.4045 | 3.551 | 1.4391 | 3.6555 | 1.3201 |
| 310 | 3.336 | 1.5027 | 3.5225 | 1.3941 | 3.5823 | 1.3459 |
| 315 | 3.322 | 1.4274 | 3.4269 | 1.4865 | 3.5416 | 1.3057 |
| 320 | 3.27 | 1.479 | 3.4116 | 1.5259 | 3.4668 | 1.4714 |
| 325 | 3.2837 | 1.4556 | 3.4011 | 1.385 | 3.4815 | 1.3729 |
| 330 | 3.2944 | 1.3648 | 3.4731 | 1.3326 | 3.4809 | 1.4419 |
| 335 | 3.2633 | 1.3718 | 3.3561 | 1.4464 | 3.5402 | 1.3025 |
| 340 | 3.2637 | 1.3718 | 3.3961 | 1.4011 | 3.494 | 1.3792 |
| 345 | 3.2641 | 1.3386 | 3.4303 | 1.3006 | 3.4961 | 1.2528 |
| 350 | 3.3023 | 1.3121 | 3.4685 | 1.2757 | 3.5294 | 1.223 |
| 355 | 3.3719 | 1.299 | 3.5294 | 1.2627 | 3.5985 | 1.2029 |
| 360 | 3.4528 | 1.3035 | 3.6136 | 1.2677 | 3.6826 | 1.1982 |
| 365 | 3.5462 | 1.3336 | 3.709 | 1.2846 | 3.7724 | 1.2182 |
| 370 | 3.6376 | 1.3899 | 3.8074 | 1.3412 | 3.8714 | 1.2563 |
| 375 | 3.7289 | 1.4511 | 3.9068 | 1.3996 | 3.9637 | 1.3177 |
| 380 | 3.8197 | 1.5418 | 4.0057 | 1.4824 | 4.0666 | 1.3915 |
| 385 | 3.8902 | 1.6507 | 4.0967 | 1.5794 | 4.1601 | 1.5025 |
| 390 | 3.9566 | 1.7793 | 4.1635 | 1.7226 | 4.2379 | 1.6201 |
| 395 | 3.988 | 1.9267 | 4.2226 | 1.8519 | 4.294 | 1.7524 |
| 400 | 4.0118 | 2.062 | 4.2658 | 1.989 | 4.3405 | 1.8796 |
| 405 | 4.0171 | 2.1822 | 4.283 | 2.1322 | 4.3609 | 2.0221 |
| 410 | 4.015 | 2.2942 | 4.2965 | 2.2563 | 4.3832 | 2.1277 |
| 415 | 4.0147 | 2.3961 | 4.3059 | 2.3725 | 4.3728 | 2.275 |
| 420 | 4.01 | 2.4927 | 4.316 | 2.4728 | 4.3648 | 2.4102 |
| 425 | 4.0238 | 2.5792 | 4.3299 | 2.5766 | 4.4053 | 2.4629 |
| 430 | 4.0706 | 2.6375 | 4.357 | 2.6694 | 4.4043 | 2.5983 |
| 435 | 4.1378 | 2.6946 | 4.4244 | 2.7221 | 4.444 | 2.6994 |
| 440 | 4.2652 | 2.6905 | 4.5255 | 2.743 | 4.5095 | 2.768 |
| 445 | 4.4052 | 2.663 | 4.6486 | 2.7324 | 4.5844 | 2.8411 |
| 450 | 4.5663 | 2.581 | 4.7981 | 2.6589 | 4.7501 | 2.7706 |
| 455 | 4.7063 | 2.4605 | 4.9346 | 2.5516 | 4.8882 | 2.6874 |
| 460 | 4.8215 | 2.315 | 5.0494 | 2.4198 | 5.0155 | 2.563 |
| 465 | 4.9044 | 2.1545 | 5.1369 | 2.2637 | 5.1158 | 2.4116 |
| 470 | 4.9532 | 1.9918 | 5.1916 | 2.1083 | 5.1808 | 2.2625 |
| 475 | 4.9761 | 1.8365 | 5.2233 | 1.9509 | 5.2231 | 2.0897 |
| 480 | 4.9771 | 1.6824 | 5.2282 | 1.7873 | 5.2387 | 1.9169 |
| 485 | 4.9572 | 1.5402 | 5.2146 | 1.6371 | 5.2336 | 1.7691 |
| 490 | 4.9225 | 1.4136 | 5.1826 | 1.5037 | 5.206 | 1.6275 |
| 495 | 4.878 | 1.3075 | 5.141 | 1.3902 | 5.1648 | 1.5082 |
| 500 | 4.8251 | 1.2151 | 5.0897 | 1.2925 | 5.1132 | 1.3959 |
| 505 | 4.7719 | 1.1405 | 5.0342 | 1.2078 | 5.0632 | 1.3087 |
| 510 | 4.7172 | 1.0777 | 4.978 | 1.1377 | 5.002 | 1.2349 |
| 515 | 4.6645 | 1.0258 | 4.9221 | 1.0764 | 4.9488 | 1.1719 |
| 520 | 4.6139 | 0.98249 | 4.8702 | 1.0307 | 4.8958 | 1.1221 |
| 525 | 4.5651 | 0.94689 | 4.8203 | 0.98996 | 4.8426 | 1.0836 |
| 530 | 4.522 | 0.91761 | 4.771 | 0.95779 | 4.7954 | 1.0463 |
| 535 | 4.4782 | 0.89218 | 4.7243 | 0.93056 | 4.7499 | 1.0169 |
| 540 | 4.4403 | 0.87335 | 4.6828 | 0.9099 | 4.707 | 0.98963 |
| 545 | 4.4041 | 0.85673 | 4.645 | 0.89235 | 4.6695 | 0.9703 |
| 550 | 4.3703 | 0.84602 | 4.6095 | 0.87971 | 4.6325 | 0.95709 |
| 555 | 4.3375 | 0.83775 | 4.576 | 0.87198 | 4.6 | 0.94654 |
| 560 | 4.3095 | 0.8319 | 4.5456 | 0.8651 | 4.5676 | 0.93708 |
| 565 | 4.283 | 0.83031 | 4.5171 | 0.8632 | 4.541 | 0.93583 |
| 570 | 4.2577 | 0.83351 | 4.4911 | 0.86544 | 4.5113 | 0.93621 |
| 575 | 4.2347 | 0.84044 | 4.4695 | 0.87351 | 4.4871 | 0.9397 |
| 580 | 4.2144 | 0.85344 | 4.4499 | 0.88519 | 4.4684 | 0.95332 |
| 585 | 4.197 | 0.87124 | 4.4335 | 0.90199 | 4.4462 | 0.96934 |
| 590 | 4.1902 | 0.89627 | 4.424 | 0.92556 | 4.4355 | 0.99475 |
| 595 | 4.1932 | 0.92731 | 4.4263 | 0.95421 | 4.4391 | 1.0275 |
| 600 | 4.2113 | 0.96158 | 4.4429 | 0.98682 | 4.4556 | 1.0629 |
| 605 | 4.2497 | 0.99109 | 4.4772 | 1.0156 | 4.4912 | 1.0969 |
| 610 | 4.3046 | 1.0072 | 4.5292 | 1.0338 | 4.5487 | 1.1162 |
| 615 | 4.3719 | 0.99878 | 4.5922 | 1.0317 | 4.6197 | 1.1127 |
| 620 | 4.4325 | 0.96306 | 4.6579 | 1.002 | 4.688 | 1.0805 |
| 625 | 4.4818 | 0.90513 | 4.7114 | 0.9479 | 4.7432 | 1.0185 |
| 630 | 4.502 | 0.8389 | 4.74 | 0.88053 | 4.7696 | 0.9466 |
| 635 | 4.4977 | 0.77679 | 4.7446 | 0.81452 | 4.7704 | 0.87832 |
| 640 | 4.4752 | 0.73226 | 4.7273 | 0.76155 | 4.7502 | 0.82588 |
| 645 | 4.4491 | 0.70943 | 4.7029 | 0.72954 | 4.7219 | 0.79517 |
| 650 | 4.432 | 0.70719 | 4.6851 | 0.7185 | 4.702 | 0.79334 |
| 655 | 4.435 | 0.72109 | 4.6814 | 0.72287 | 4.699 | 0.80448 |
| 660 | 4.467 | 0.73742 | 4.704 | 0.73177 | 4.729 | 0.82186 |
| 665 | 4.5298 | 0.73315 | 4.7528 | 0.72707 | 4.7919 | 0.82357 |
| 670 | 4.6078 | 0.68726 | 4.8171 | 0.68928 | 4.8726 | 0.78365 |
| 675 | 4.666 | 0.59454 | 4.8704 | 0.61454 | 4.9433 | 0.69605 |
| 680 | 4.6761 | 0.48363 | 4.8903 | 0.51437 | 4.963 | 0.5781 |
| 685 | 4.647 | 0.38463 | 4.8732 | 0.41919 | 4.943 | 0.46674 |
| 690 | 4.598 | 0.3088 | 4.8323 | 0.33856 | 4.8923 | 0.38264 |
| 695 | 4.5437 | 0.25494 | 4.7814 | 0.2786 | 4.8365 | 0.31781 |
| 700 | 4.4906 | 0.21878 | 4.7296 | 0.23671 | 4.7787 | 0.27401 |
| 705 | 4.442 | 0.19383 | 4.6795 | 0.20579 | 4.7266 | 0.24176 |
| 710 | 4.3986 | 0.1754 | 4.6358 | 0.18508 | 4.6804 | 0.21788 |
| 715 | 4.3601 | 0.16272 | 4.5942 | 0.1686 | 4.6386 | 0.20133 |
| 720 | 4.3252 | 0.15274 | 4.5574 | 0.15747 | 4.5989 | 0.19057 |
| 725 | 4.2952 | 0.14619 | 4.5273 | 0.14864 | 4.5628 | 0.18154 |
| 730 | 4.2678 | 0.1395 | 4.4969 | 0.14141 | 4.5366 | 0.1743 |
| 735 | 4.2421 | 0.13593 | 4.4708 | 0.13618 | 4.5081 | 0.16678 |
| 740 | 4.217 | 0.13169 | 4.4458 | 0.13096 | 4.481 | 0.16298 |
| 745 | 4.1952 | 0.12704 | 4.4218 | 0.12653 | 4.4563 | 0.15832 |
| 750 | 4.1743 | 0.12604 | 4.4002 | 0.12364 | 4.4345 | 0.15283 |
| 755 | 4.1555 | 0.12129 | 4.3803 | 0.12012 | 4.4136 | 0.15051 |
| 760 | 4.1389 | 0.1201 | 4.3617 | 0.1169 | 4.3922 | 0.14691 |
| 765 | 4.1227 | 0.11758 | 4.3437 | 0.11588 | 4.3773 | 0.14221 |
| 770 | 4.1064 | 0.11639 | 4.3282 | 0.11193 | 4.3573 | 0.14208 |
| 775 | 4.0927 | 0.11393 | 4.3129 | 0.11032 | 4.3438 | 0.13904 |
| 780 | 4.0796 | 0.1128 | 4.2978 | 0.11047 | 4.3291 | 0.1358 |
| 785 | 4.0668 | 0.11275 | 4.2846 | 0.1064 | 4.3149 | 0.13477 |
| 790 | 4.0534 | 0.11062 | 4.2707 | 0.10698 | 4.3012 | 0.13296 |
| 795 | 4.0397 | 0.11007 | 4.2593 | 0.10523 | 4.286 | 0.12998 |
| 800 | 4.0301 | 0.10719 | 4.2458 | 0.10259 | 4.2751 | 0.12882 |
| 805 | 4.0179 | 0.10728 | 4.2331 | 0.10131 | 4.2621 | 0.12699 |
| 810 | 4.0077 | 0.10608 | 4.2227 | 0.099294 | 4.2491 | 0.12501 |
| 815 | 3.9976 | 0.10568 | 4.2115 | 0.097373 | 4.2399 | 0.12492 |
| 820 | 3.9891 | 0.10431 | 4.2017 | 0.096565 | 4.2289 | 0.12063 |
| 825 | 3.9809 | 0.10323 | 4.1931 | 0.095439 | 4.22 | 0.12048 |
| 830 | 3.9721 | 0.10319 | 4.1842 | 0.095125 | 4.2102 | 0.11856 |
| 835 | 3.9637 | 0.10154 | 4.1749 | 0.093883 | 4.1994 | 0.11879 |
| 840 | 3.955 | 0.10175 | 4.1656 | 0.093628 | 4.1892 | 0.11732 |
| 845 | 3.9489 | 0.10215 | 4.1581 | 0.092538 | 4.1822 | 0.11646 |
| 850 | 3.9379 | 0.09965 | 4.15 | 0.091605 | 4.1743 | 0.11646 |
| 855 | 3.9318 | 0.10056 | 4.1397 | 0.091592 | 4.164 | 0.11486 |
| 860 | 3.9243 | 0.09773 | 4.1329 | 0.091321 | 4.1572 | 0.1142 |
| 865 | 3.9194 | 0.097773 | 4.1253 | 0.088939 | 4.1489 | 0.11333 |
| 870 | 3.9123 | 0.096894 | 4.118 | 0.089321 | 4.1426 | 0.11253 |
| 875 | 3.906 | 0.096658 | 4.1108 | 0.086885 | 4.1345 | 0.11163 |
| 880 | 3.8986 | 0.095618 | 4.1026 | 0.086427 | 4.1266 | 0.10953 |
| 885 | 3.892 | 0.09492 | 4.097 | 0.086398 | 4.1205 | 0.10868 |
| 890 | 3.8872 | 0.095464 | 4.0899 | 0.086771 | 4.1144 | 0.10922 |
| 895 | 3.8807 | 0.093879 | 4.0835 | 0.08629 | 4.1077 | 0.1072 |
| 900 | 3.8756 | 0.093993 | 4.0769 | 0.085472 | 4.1013 | 0.10737 |
| 905 | 3.8676 | 0.092612 | 4.0717 | 0.085391 | 4.0954 | 0.10671 |
| 910 | 3.8641 | 0.094722 | 4.0659 | 0.085023 | 4.0873 | 0.10872 |
| 915 | 3.8589 | 0.092782 | 4.0599 | 0.085014 | 4.0827 | 0.10507 |
| 920 | 3.8547 | 0.093257 | 4.056 | 0.084889 | 4.0774 | 0.10645 |
| 925 | 3.85 | 0.09313 | 4.0507 | 0.084378 | 4.0728 | 0.10621 |
| 930 | 3.8441 | 0.092887 | 4.0461 | 0.085745 | 4.0668 | 0.10531 |
| 935 | 3.8401 | 0.091613 | 4.0421 | 0.084569 | 4.0631 | 0.10413 |
| 940 | 3.8355 | 0.092333 | 4.0365 | 0.083004 | 4.0568 | 0.10517 |
| 945 | 3.8314 | 0.092179 | 4.0302 | 0.084009 | 4.0512 | 0.10484 |
| 950 | 3.826 | 0.090759 | 4.0249 | 0.082722 | 4.0484 | 0.10433 |
| 955 | 3.8218 | 0.09071 | 4.0196 | 0.082297 | 4.0429 | 0.10295 |
| 960 | 3.8184 | 0.09035 | 4.0163 | 0.083586 | 4.0382 | 0.10119 |
| 965 | 3.8119 | 0.090305 | 4.0132 | 0.082068 | 4.034 | 0.10269 |
| 970 | 3.8114 | 0.089903 | 4.0075 | 0.08227 | 4.0285 | 0.10053 |
| 975 | 3.8056 | 0.091299 | 4.003 | 0.082816 | 4.0233 | 0.1022 |
| 980 | 3.8006 | 0.091263 | 3.9987 | 0.082014 | 4.0206 | 0.10192 |
| 985 | 3.798 | 0.090927 | 3.9955 | 0.083107 | 4.0154 | 0.10174 |
| 990 | 3.7938 | 0.091766 | 3.9911 | 0.08252 | 4.0111 | 0.10077 |
| 995 | 3.7913 | 0.091764 | 3.9878 | 0.082941 | 4.0073 | 0.1008 |
| 1000 | 3.7867 | 0.093497 | 3.9844 | 0.082239 | 4.0022 | 0.10074 |

## 10L

| wavelength(nm) | n | k |
|---|---|---|
| 275 | 3.8889 | 0.46559 |
| 280 | 3.9301 | 0.83355 |
| 285 | 3.8389 | 0.86449 |
| 290 | 3.8264 | 1.0317 |
| 295 | 3.7683 | 1.0961 |
| 300 | 3.7541 | 1.2344 |
| 305 | 3.7127 | 1.2669 |
| 310 | 3.6691 | 1.252 |
| 315 | 3.6259 | 1.275 |
| 320 | 3.5782 | 1.345 |
| 325 | 3.5906 | 1.2531 |
| 330 | 3.5811 | 1.2913 |
| 335 | 3.5242 | 1.3573 |
| 340 | 3.5243 | 1.3895 |
| 345 | 3.5694 | 1.2122 |
| 350 | 3.6053 | 1.1739 |
| 355 | 3.6695 | 1.1487 |
| 360 | 3.7566 | 1.148 |
| 365 | 3.8485 | 1.157 |
| 370 | 3.9498 | 1.2021 |
| 375 | 4.0538 | 1.2517 |
| 380 | 4.1583 | 1.3426 |
| 385 | 4.2525 | 1.4359 |
| 390 | 4.3427 | 1.5527 |
| 395 | 4.4126 | 1.6849 |
| 400 | 4.4668 | 1.8126 |
| 405 | 4.4984 | 1.9556 |
| 410 | 4.5164 | 2.0879 |
| 415 | 4.5229 | 2.2223 |
| 420 | 4.5349 | 2.3319 |
| 425 | 4.5421 | 2.4486 |
| 430 | 4.5628 | 2.5578 |
| 435 | 4.6021 | 2.6651 |
| 440 | 4.6889 | 2.7158 |
| 445 | 4.7219 | 2.8865 |
| 450 | 4.8609 | 2.8627 |
| 455 | 5.0147 | 2.798 |
| 460 | 5.1515 | 2.6832 |
| 465 | 5.2829 | 2.4814 |
| 470 | 5.3587 | 2.2918 |
| 475 | 5.4033 | 2.1126 |
| 480 | 5.4136 | 1.9448 |
| 485 | 5.4078 | 1.7579 |
| 490 | 5.3769 | 1.6211 |
| 495 | 5.33 | 1.4952 |
| 500 | 5.2755 | 1.377 |
| 505 | 5.2115 | 1.2789 |
| 510 | 5.1501 | 1.2042 |
| 515 | 5.0907 | 1.1464 |
| 520 | 5.032 | 1.0924 |
| 525 | 4.976 | 1.0438 |
| 530 | 4.9239 | 1.0179 |
| 535 | 4.8733 | 0.98939 |
| 540 | 4.8288 | 0.95548 |
| 545 | 4.7877 | 0.94788 |
| 550 | 4.7497 | 0.93046 |
| 555 | 4.7173 | 0.91882 |
| 560 | 4.6814 | 0.91307 |
| 565 | 4.6495 | 0.90966 |
| 570 | 4.6224 | 0.90747 |
| 575 | 4.593 | 0.91041 |
| 580 | 4.5699 | 0.92088 |
| 585 | 4.549 | 0.94077 |
| 590 | 4.5356 | 0.97007 |
| 595 | 4.536 | 1.0036 |
| 600 | 4.5524 | 1.0479 |
| 605 | 4.59 | 1.0794 |
| 610 | 4.6502 | 1.1032 |
| 615 | 4.7286 | 1.1016 |
| 620 | 4.8042 | 1.0707 |
| 625 | 4.8661 | 1.0051 |
| 630 | 4.8941 | 0.9276 |
| 635 | 4.8926 | 0.84516 |
| 640 | 4.8683 | 0.78793 |
| 645 | 4.8297 | 0.75324 |
| 650 | 4.8004 | 0.75122 |
| 655 | 4.7931 | 0.77326 |
| 660 | 4.8199 | 0.80189 |
| 665 | 4.8913 | 0.81809 |
| 670 | 4.9919 | 0.78371 |
| 675 | 5.0777 | 0.68249 |
| 680 | 5.1046 | 0.55076 |
| 685 | 5.0776 | 0.42479 |
| 690 | 5.0175 | 0.32756 |
| 695 | 4.9536 | 0.26415 |
| 700 | 4.891 | 0.21591 |
| 705 | 4.837 | 0.18612 |
| 710 | 4.7828 | 0.16645 |
| 715 | 4.7347 | 0.14724 |
| 720 | 4.6958 | 0.13861 |
| 725 | 4.6612 | 0.12378 |
| 730 | 4.6303 | 0.12181 |
| 735 | 4.5992 | 0.11637 |
| 740 | 4.5678 | 0.11279 |
| 745 | 4.5464 | 0.1092 |
| 750 | 4.5231 | 0.10622 |
| 755 | 4.4982 | 0.10033 |
| 760 | 4.4776 | 0.10159 |
| 765 | 4.46 | 0.095861 |
| 770 | 4.4429 | 0.096327 |
| 775 | 4.4271 | 0.092949 |
| 780 | 4.4096 | 0.092671 |
| 785 | 4.3944 | 0.0904 |
| 790 | 4.3814 | 0.089364 |
| 795 | 4.3657 | 0.089596 |
| 800 | 4.3543 | 0.085195 |
| 805 | 4.3403 | 0.086865 |
| 810 | 4.3274 | 0.08522 |
| 815 | 4.3174 | 0.081597 |
| 820 | 4.3074 | 0.080824 |
| 825 | 4.2978 | 0.081814 |
| 830 | 4.2874 | 0.080366 |
| 835 | 4.2757 | 0.077703 |
| 840 | 4.2655 | 0.078495 |
| 845 | 4.2572 | 0.081157 |
| 850 | 4.2466 | 0.077832 |
| 855 | 4.2395 | 0.079486 |
| 860 | 4.2321 | 0.078655 |
| 865 | 4.2248 | 0.075373 |
| 870 | 4.2151 | 0.07729 |
| 875 | 4.2097 | 0.074161 |
| 880 | 4.2004 | 0.071378 |
| 885 | 4.1926 | 0.072436 |
| 890 | 4.1867 | 0.073229 |
| 895 | 4.1784 | 0.071606 |
| 900 | 4.1725 | 0.074564 |
| 905 | 4.1657 | 0.07173 |
| 910 | 4.158 | 0.070905 |
| 915 | 4.1541 | 0.07066 |
| 920 | 4.1469 | 0.071455 |
| 925 | 4.1423 | 0.070668 |
| 930 | 4.1391 | 0.070459 |
| 935 | 4.1314 | 0.071227 |
| 940 | 4.1272 | 0.070284 |
| 945 | 4.1227 | 0.070336 |
| 950 | 4.117 | 0.068629 |
| 955 | 4.1106 | 0.068798 |
| 960 | 4.1089 | 0.070415 |
| 965 | 4.1046 | 0.069971 |
| 970 | 4.0976 | 0.070051 |
| 975 | 4.0931 | 0.068335 |
| 980 | 4.0894 | 0.067586 |
| 985 | 4.0849 | 0.068757 |
| 990 | 4.0802 | 0.067641 |
| 995 | 4.0755 | 0.067518 |
| 1000 | 4.0732 | 0.068066 |

# References

S1  Yu, Y. F. *et al.* Controlled Scalable Synthesis of Uniform, High-Quality Monolayer and Few-layer MoS2 Films. *Sci Rep-Uk* **3**, 1866 (2013).

S2  Kresse, G. & Furthmuller, J. Efficient iterative schemes for ab initio total-energy calculations using a plane-wave basis set. *Physical Review B* **54**, 11169-11186, (1996).

S3  Becke, A. D. Density-Functional Exchange-Energy Approximation with Correct Asymptotic-Behavior. *Phys Rev A* **38**, 3098-3100, (1988).

S4  Shi, H. L., Pan, H., Zhang, Y. W. & Yakobson, B. I. Quasiparticle band structures and optical properties of strained monolayer MoS2 and WS2. *Physical Review B* **87**, 155304 (2013).